\documentclass[twocolumn,reprint,amsmath,superscriptaddress,amssymb,aps,a4paper]{revtex4-2}
\usepackage[utf8]{inputenc}
\usepackage{minitoc}
\usepackage{caption}
\usepackage{graphicx}
\usepackage{subcaption}
\usepackage{dcolumn}
\usepackage{bm}
\usepackage{siunitx}
\usepackage{booktabs}
\usepackage[usenames,dvipsnames]{xcolor}
\usepackage{tcolorbox}
\usepackage{tabularx}
\usepackage{array}
\usepackage{colortbl}
\usepackage{braket}
\usepackage{gensymb}
\usepackage{pdfpages}
\usepackage{amsmath}
\usepackage{mathrsfs}
\usepackage{blindtext}
\usepackage{minitoc}
\usepackage{etoolbox}
\usepackage{float}
\usepackage{physics}

\usepackage[colorlinks=true, citecolor=blue, linkcolor=blue, urlcolor=blue]{hyperref}

\usepackage{xfrac}
\usepackage{geometry}
\usepackage{bbold}
\usepackage{xcolor}
\usepackage{color,soul}

\usepackage{ragged2e}
\makeatletter
\long\def\@makecaption#1#2{%
  \vskip\abovecaptionskip
  \sbox\@tempboxa{\small\justifying #1: #2}%
  \ifdim \wd\@tempboxa > \columnwidth
    \small\justifying #1: #2\par
  \else
    \global \@minipagefalse
    \hb@xt@\columnwidth{\hfil\box\@tempboxa\hfil}%
  \fi
  \vskip\belowcaptionskip}
\makeatother

\makeatletter
\patchcmd{\@outputpage@head}{\@ifx{\LS@rot\@undefined}{}{\LS@rot}}{}{}{}
\makeatother

\tcbuselibrary{skins}

\newcommand{\units}[1]{\,\mathrm{#1}}
\tcbset{tab2/.style={colback=gray!5!white,colframe=black!50!black,colbacktitle=Gray!40!white,
coltitle=black,center title}}

\begin{document}

\title{Quantum sensing with a spin ensemble in a van der Waals material}

\author{Souvik Biswas}
\thanks{These authors contributed equally to this work.}
\affiliation{Department of Electrical Engineering, Stanford University, Stanford, CA, USA}

\author{Giovanni Scuri}
\thanks{These authors contributed equally to this work.}
\affiliation{Department of Electrical Engineering, Stanford University, Stanford, CA, USA}

\author{Noah Huffman}
\affiliation{Department of Physics, Stanford University, Stanford, CA, USA}

\author{Eric I. Rosenthal}
\affiliation{Department of Electrical Engineering, Stanford University, Stanford, CA, USA}
\affiliation{Present address: Sygaldry Technologies, Ann Arbor MI, USA}

\author{Ruotian Gong}
\affiliation{Department of Physics, Washington University, St. Louis, MO, USA}

\author{Thomas Poirier}
\affiliation{Tim Taylor Department of Chemical Engineering, Kansas State University, Manhattan, KS, USA}

\author{Xingyu Gao}
\affiliation{Department of Physics and Astronomy, Purdue University, West Lafayette, IN, USA}

\author{Sumukh Vaidya}
\affiliation{Department of Physics and Astronomy, Purdue University, West Lafayette, IN, USA}

\author{Abigail J. Stein}
\affiliation{Department of Applied Physics, Stanford University, Stanford, CA, USA}

\author{Tsachy Weissman}
\affiliation{Department of Electrical Engineering, Stanford University, Stanford, CA, USA}

\author{James H. Edgar}
\affiliation{Tim Taylor Department of Chemical Engineering, Kansas State University, Manhattan, KS, USA}

\author{Tongcang Li}
\affiliation{Department of Physics and Astronomy, Purdue University, West Lafayette, IN, USA}
\affiliation{Elmore Family School of Electrical and Computer Engineering, Purdue University, West Lafayette, IN, USA}

\author{Chong Zu}
\affiliation{Department of Physics, Washington University, St. Louis, MO, USA}

\author{Jelena Vu\v{c}kovi\'{c}}
\email{jela@stanford.edu}
\affiliation{Department of Electrical Engineering, Stanford University, Stanford, CA, USA}

\author{Joonhee Choi}
\email{joonhee.choi@stanford.edu}
\affiliation{Department of Electrical Engineering, Stanford University, Stanford, CA, USA}

\date{\today}

\begin{abstract}
Quantum sensing with solid-state spin defects has transformed nanoscale metrology, offering subwavelength spatial resolution with exceptional sensitivity to multiple signal types~\cite{degen2017quantum, aslam2023quantum, du2024single, rovny2024nanoscale}. Maximizing these advantages requires minimizing both the sensor-target separation and detectable signal threshold. However, leading platforms such as nitrogen-vacancy (NV) centers in diamond suffer from performance degradation near surfaces~\cite{kim2015decoherence, sangtawesin2019origins} or in nanoscale volumes~\cite{mcguinness2011quantum, holzgrafe2020nanoscale, miller2020spin}, motivating the search for optically addressable spin sensors in atomically thin, two-dimensional (2D) van der Waals materials~\cite{lee2022spin, clua2023isotopic, gong2024isotope, stern2022room, stern2024quantum, gao2025single, gottscholl2021spin, gottscholl2021room, sasaki2023magnetic, rizzato2023extending}. Here, we present a comprehensive experimental framework to probe a 2D spin ensemble, including its Hamiltonian, coherent sensing dynamics, and noise environment. Using a central spin system in a hexagonal boron nitride (hBN) crystal, we fully map the hyperfine interactions with proximal nuclear spins, demonstrate switchable magnetic and electric noise sensing, and introduce a method to accurately reconstruct the environmental noise spectrum explicitly accounting for quantum control imperfections. We achieve a record coherence time of 80~$\mu$s under dynamical decoupling, enabling sub-microtesla AC magnetic sensitivity at a 10 nm target distance. Leveraging the broad opportunities for defect engineering in atomically thin hosts~\cite{ye2019spin, vaidya2023quantum, fang2024quantum}, these results lay the foundation for next-generation quantum sensors with ultrahigh sensitivity, tunable noise selectivity, and versatile functionalities~\cite{aslam2023quantum, du2024single, rovny2024nanoscale}.
\end{abstract}

\maketitle

\begin{figure*}[htbp] 
\centering
\includegraphics[scale=1]{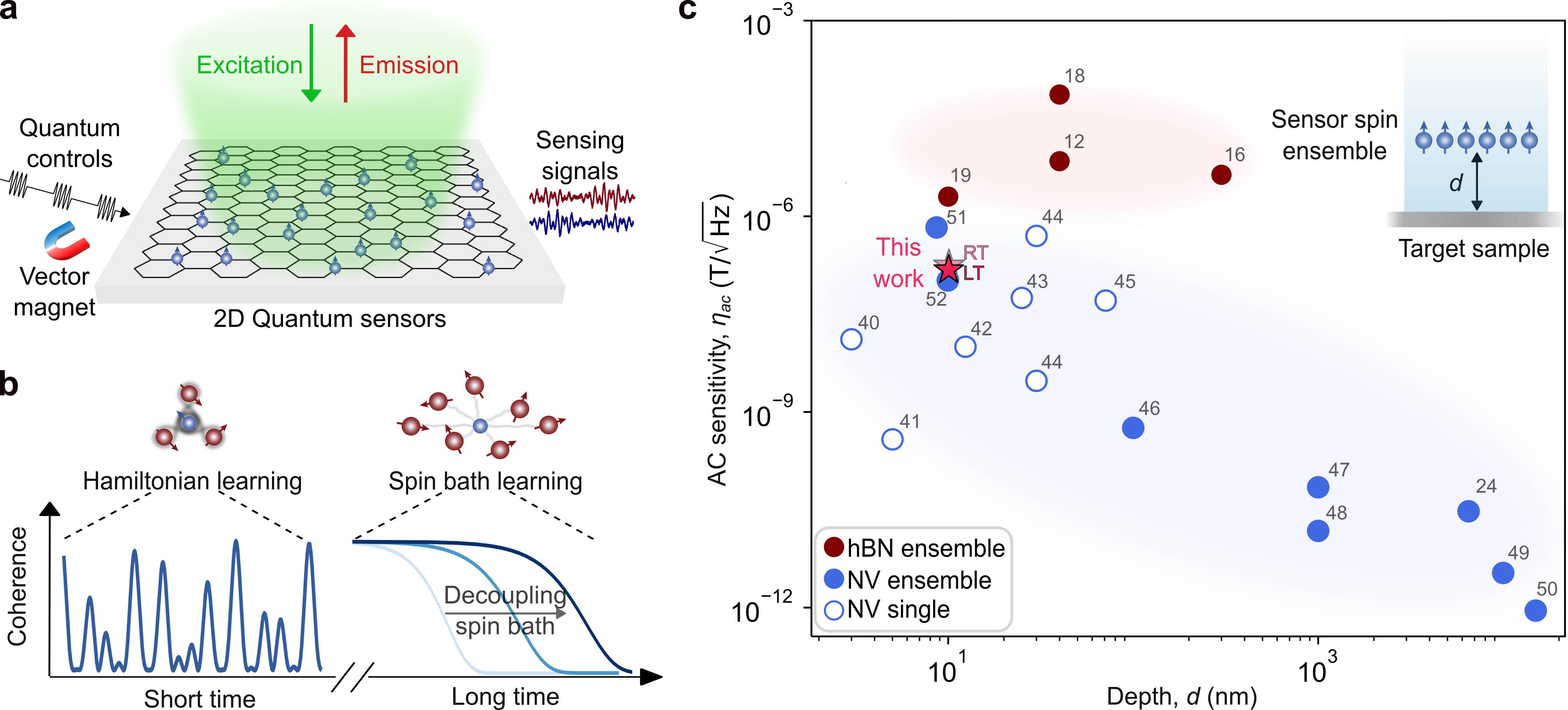}
\caption{\textbf{Experimental framework and sensing advantages of 2D quantum sensor platforms.} \textbf{a,} Optically addressable spin defects in 2D materials (blue spins) provide a versatile sensing platform for detecting weak environmental signals, with programmable performance enabled by a vector magnetic field and robust quantum controls. \textbf{b,} Such spin-defect sensors are typically surrounded by nuclear spins in the bath (red spins), forming a central spin system. Monitoring the sensor’s coherence under various control sequences enables probing of coherent many-body dynamics at short times, and of the spin bath and external noise at longer times. \textbf{c,} 2D sensor platforms offer advantages for nanoscale sensing applications due to their close proximity to the target sample (inset). To illustrate this, we plot and compare reported AC magnetic field sensitivities, $\eta_\text{ac}$, as a function of the sensor-sample distance, $d$, for NV centers in diamond and boron-vacancy ensemble sensors, including our work at both room temperature ($\approx300\units{K}$) and low temperature ($\approx2\units{K}$). The achieved sensitivities (red stars) are already comparable to state-of-the-art systems. The numbers next to the markers indicate the corresponding references. We assume that the target signal is globally correlated across the ensemble to facilitate comparison between single and ensemble sensors. }
\label{fig:Fig1}
\end{figure*}


Solid-state spin systems have emerged as powerful sensing platforms, allowing access to unprecedented regimes in nanoscale magnetic resonance imaging~\cite{aslam2017nanoscale, glenn2018high}, single-molecule spectroscopy~\cite{shi2015single, lovchinsky2016nuclear}, local probing of condensed matter systems~\cite{andersen2019electron, esat2024quantum}, electromagnetic and temperature sensing in living cells~\cite{kucsko2013nanometre, choi2020probing}, and studies of magnetism in geological samples~\cite{fu2014solar, glenn2017micrometer}. Leveraging rapid progress in the synthesis and control of 2D materials~\cite{geim2013van, castellanos2022van}, the exploration of optically bright and spin-coherent defects in van der Waals hosts has recently attracted considerable interest~\cite{stern2022room, clua2023isotopic, gong2024isotope, stern2024quantum, gao2025single, gottscholl2021spin, gottscholl2021room, sasaki2023magnetic, rizzato2023extending}, as their pristine crystallographic structures and versatile engineering potential provide promising routes to preserve quantum coherence while maintaining nanometric spatial resolution.

Several promising defects -- such as carbon-related centers~\cite{stern2022room, stern2024quantum, gao2025single} and boron vacancies in isotopically purified hBN~\cite{clua2023isotopic, gong2024isotope} -- have recently been reported in van der Waals materials. Despite microscopic differences in crystallographic structure, most of these quantum sensors can be broadly viewed as {\it central spin systems}~\cite{onizhuk2025colloquium}, where a single optically accessible spin is coupled to a number of optically dark spins in the environment. However, a key challenge is to resolve both the effective Hamiltonian that governs hyperfine interactions~\cite{rao2016characterization, abobeih2019atomic} and the noise processes that limit sensor coherence~\cite{bauch2020decoherence, wise2021using}. Meeting this challenge calls for device-agnostic, experimentally efficient protocols that fully unlock the sensing potential of spin defects in engineered 2D materials.

To this end, we present a framework for probing and engineering the quantum dynamics of a central spin system in a 2D material, using a tunable external vector magnet and robust quantum controls (Fig.~\ref{fig:Fig1}a). In particular, applying a strong magnetic field orthogonal to the central spin’s quantization axis enables precise control over its susceptibility to various noise sources. This programmable sensitivity can be leveraged to gain rich insights into coherent quantum interactions with neighboring nuclear spins at short times as well as couplings to both the spin bath and extrinsic noise sources at longer times---revealed through the sensor’s time-domain decoherence dynamics (Fig.~\ref{fig:Fig1}b). Remarkably, we find that the AC magnetic field sensitivity of our 2D ensemble quantum sensor, based on boron vacancies in hBN \cite{rizzato2023extending, sasaki2023magnetic, gong2024isotope, gottscholl2021spin}, already rivals that of state-of-the-art platforms such as NV centers in diamond \cite{zheng2022coherence, rosskopf2014investigation, myers2014probing, hong2013nanoscale, maze2008nanoscale, palm2022imaging, chen2022nanodiamond, shim2022multiplexed, barry2016optical, glenn2018high, arunkumar2023quantum, wolf2015subpicotesla, Ziem2019WideFieldNV, Tetienne2018DenseNV}, which typically require much more complex surface treatment and passivation (Fig.~\ref{fig:Fig1}c). Our results highlight the promising future and immediate utility of 2D-material-based quantum sensors for nanoscale sensing with unprecedented precision. 

\begin{figure*}[htbp] 
\begin{center}
\includegraphics[scale=1]{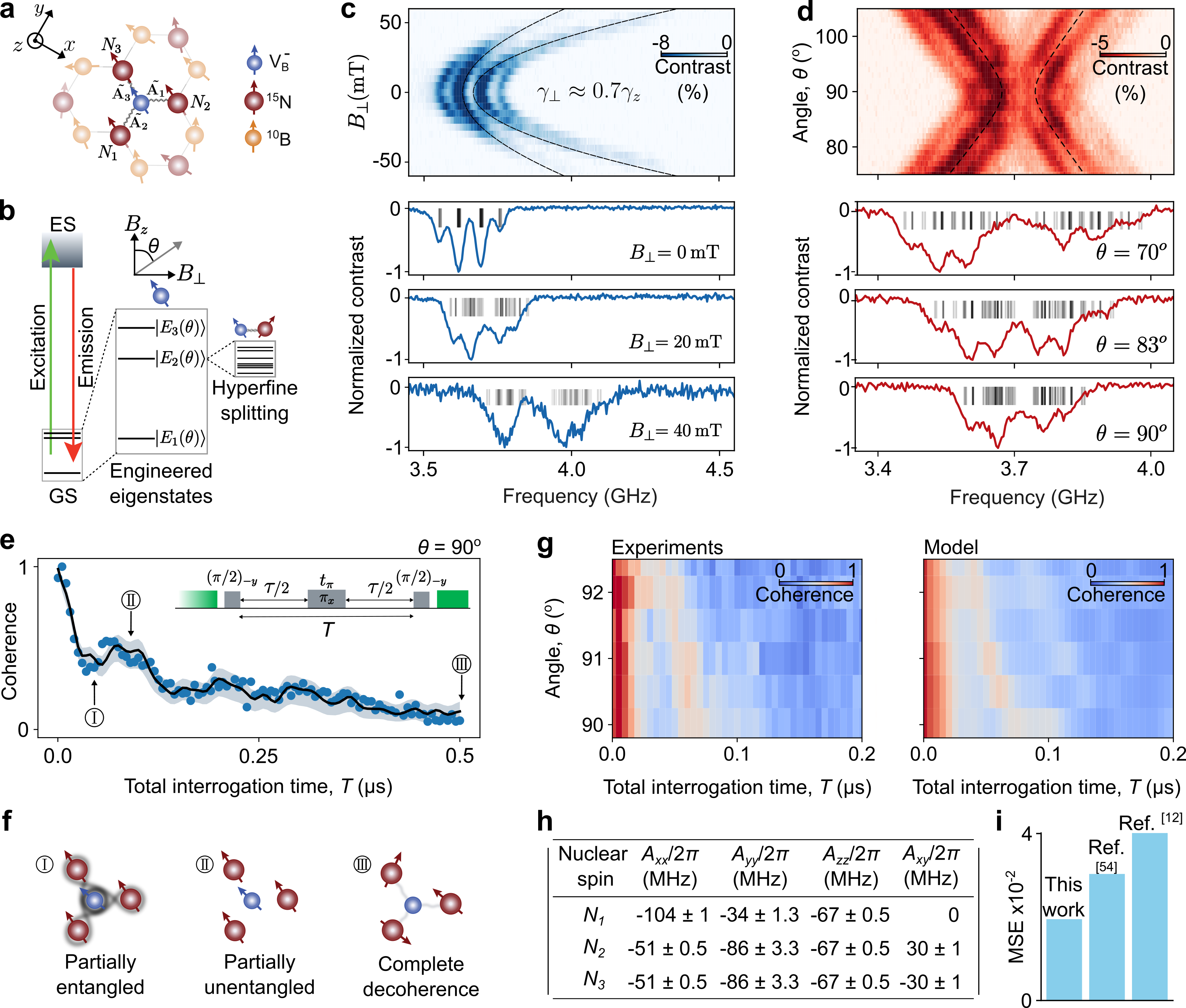}
\caption{\textbf{Probing electron-nuclear hyperfine interactions in a central spin system.} \textbf{a,} An electronic spin from a boron-vacancy defect, $V_B^-$ (blue spin), in hBN couples strongly to each $i$-th nearest-neighbor $^{15}$N nuclear spin, $N_i$ (red), via coherent hyperfine interaction, $\tilde{A}_i$, and more weakly to the surrounding spin bath of nitrogen and boron ($^{10}$B, orange). \textbf{b,} The central spin in the ground-state (GS) manifold forms a spin-1 system with eigenstates, $\{\ket{E_1(\theta)}, \ket{E_2(\theta)}, \ket{E_3(\theta)}\}$, which can be engineered by applying a magnetic field at an angle $\theta$ relative to the $z$-axis. Hyperfine splittings depend on the engineered states. The GS levels are probed via ODMR spectroscopy involving spin-dependent excited-state (ES) dynamics (Methods)~\cite{lee2025intrinsic}. \textbf{c,} ODMR spectra measured as a function of in-plane magnetic field strength, $B_\perp$, at $\theta = 90^\circ$ show that a simple four-peak structure broadens into widely distributed resonances as $B_\perp$ increases. \textbf{d,} Rotating the external field near the in-plane direction ($\theta = 90^\circ$) at a fixed $B_\perp = 20$~mT reveals complex spectral variations. In \textbf{c} and \textbf{d}, the black barcode lines mark the theoretically expected resonance positions based on the learned hyperfine parameters (see Table in \textbf{h}). We find that the in-plane gyromagnetic ratio, $\gamma_\perp$, is $\approx$~30\% smaller than the out-of-plane gyromagnetic ratio, $\gamma_z$ (black dashed lines). \textbf{e,} Short-time decoherence dynamics of the central spin under a spin-echo protocol (inset) enable extraction of the hyperfine parameters by fitting the measured coherence profile (blue markers) to the analytical model (black line). The gray shading denotes error bars. \textbf{f,} 
Physically, the coherence modulation at time \textcircled{\tiny I} corresponds to partial entanglement with the nearest-neighbor nuclear spins. The modulation at time \textcircled{\tiny II} corresponds to disentanglement with the same nearest-neighbor spins. Finally, time \textcircled{\tiny III} represents complete decoherence arising from coupling to the larger bath. \textbf{g,} Sensitive $\theta$-dependent spin-echo dynamics enable calibration of the hyperfine parameters, as summarized in \textbf{h}, yielding excellent agreement between model and experiment. \textbf{i,} Benchmarking against density functional theory (DFT) predictions of Ref.~\cite{ivady2020ab} and experimental work of Ref.~\cite{gong2024isotope} shows that our parameters provide the best fit with the lowest mean-square errors (MSE) (Extended Data Fig.~\ref{fig:compare}).}
\label{fig:Fig2}
\end{center}
\end{figure*}

\subsection*{Probing and Characterizing a 2D Quantum Sensor}

Our quantum sensing platform consists of an ensemble of negatively charged boron vacancies, $V_B^-$, implanted into a 2D hBN lattice (Extended Data Fig.~\ref{fig:setup}). While our system consists of multiple hBN layers with a total thickness of $\approx$10 nm, we approximate it as a set of uncoupled stacked 2D lattices, given the boron vacancy density and inter-defect interaction strength (Methods). Within the crystalline host, the $V_B^-$ defect is surrounded by boron and nitrogen atoms, both of which possess nuclear spins. The nitrogen atoms in our sample are isotopically engineered to $^{15}$N with a nuclear spin of $1/2$ (Methods)~\cite{gong2024isotope, clua2023isotopic}. Such a single defect, surrounded by many nuclear spins in the bath, is prototypical of a central spin system. In particular, the three nearest-neighbor nitrogen atoms interact independently with the central $V_B^-$ defect via the hyperfine interaction (Fig.~\ref{fig:Fig2}a). 

Similar to NV centers in diamond, the boron-vacancy center is an optically addressable defect whose ground-state (GS) manifold comprises an effective spin-1 system with three magnetic sublevels, $\{\ket{0}, \ket{-1}, \ket{+1}\}$, where the nearly degenerate $\ket{\pm 1}$ states are split from the $\ket{0}$ state by a quantizing, zero-field splitting (ZFS) of $D/2\pi \approx 3.65$~GHz (Fig.~\ref{fig:Fig2}b)~\cite{liu2025temperature}. The spin population in the GS manifold can be efficiently polarized via optical pumping, manipulated with microwave driving, and read out through spin-state-dependent photoluminescence detection (Extended Data Fig.~\ref{fig:rabi}, Methods)~\cite{lee2025intrinsic}.

Importantly, by applying an external magnetic field at an angle $\theta$ relative to the surface normal of the lattice, the bare spin-1 eigenstates can be transformed into a tunable basis parameterized by $\theta$: $\{\ket{E_1(\theta)}, \ket{E_2(\theta)}, \ket{E_3(\theta)}\}$ (Fig.~\ref{fig:Fig2}b). Crucially, these engineered eigenstates exhibit $\theta$-dependent sensitivities to distinct components of the hyperfine Hamiltonian as well as to different types of external noise, providing a systematic means to uncover the microscopic mechanisms that govern both coherent and incoherent interactions in a central spin system.

Specifically, the effective Hamiltonian of our spin system, $\hat{H}$, is given by $\hat{H} = \hat{H}_0 + \hat{H}_\text{HF}$, where $\hat{H}_0$ is the dominant Hamiltonian of the central boron-vacancy sensor and $\hat{H}_\text{HF}$ is the perturbative Hamiltonian describing hyperfine coupling to the three nearest-neighbor nuclear spins (Methods). Longer-range, weaker interactions with more distant nuclear spins are treated as detuning noise, modeled as inhomogeneous broadening. The perturbative hyperfine Hamiltonian takes the form ($\hbar = 1$):
\begin{align}
    \hat{H}_\text{HF} =  \sum_{i=1}^3 \sum_{\mu,\nu =x,y,z} A^{(i)}_{\mu\nu} \hat{S}_{\mu} \hat{I}^{(i)}_{\nu}, \label{eq:H_HF}
\end{align} 
where $A^{(i)}_{\mu\nu}$ is the $(\mu,\nu)$ component of the hyperfine tensor, $\tilde{A}_i$, for the $i$-th nitrogen nuclear spin coupled to the boron vacancy’s electronic spin, $\hat{I}^{(i)}_{\nu}$ are the nuclear spin-1/2 operators, and $\hat{S}_{\mu}$ are the electronic spin-1 operators of the boron vacancy, with $\mu,\nu = x,y,z$ (Methods). 

Learning the hyperfine Hamiltonian is equivalent to extracting all tensor components of $A^{(i)}_{\mu\nu}$. In our platform, the crystal symmetry of 2D hBN enforces $\tilde{A}_2 = R_{120^\circ} \tilde{A}_1 R_{120^\circ}^T$ and $\tilde{A}_3 = R_{240^\circ} \tilde{A}_1 R_{240^\circ}^T$, where $R_\xi$ denotes an in-plane rotation about the $z$-axis by angle $\xi$~\cite{ivady2020ab}. Moreover, all three sites share a common axial strength $A_{zz} = A^{(i)}_{zz}$ and satisfy $A^{(i)}_{xz} = A^{(i)}_{yz} = 0$ for $i = 1,2,3$, thus substantially reducing the number of free parameters that must be determined (Methods).

To experimentally extract the hyperfine parameters, we first perform optically detected magnetic resonance (ODMR) spectroscopy under a tunable vector magnetic field (Figs.~\ref{fig:Fig2}c,d). ODMR enables measurement of the transition frequencies of the central sensor, perturbed by hyperfine coupling. Specifically, in the absence of an external field, the bare sensor eigenstates are the magnetic Zeeman sublevels, $\{ \ket{E_1} = \ket{0}, \ket{E_2} = \ket{-1}, \ket{E_3} = \ket{+1} \}$, with nearly-degenerate resonance frequencies set by the ZFS (Methods). The hyperfine interactions perturb these transitions, giving rise to four distinct peaks primarily determined by the axial $A_{zz}$ component and local strain, $\mathcal{E}$ (see the $B_\perp = 0$ subplot in Fig.~\ref{fig:Fig2}c, Methods). From the spectral separations between these peaks, we extract $A_{zz}/2\pi \approx -67$~MHz and $\mathcal{E}/2\pi \approx 18.5$~MHz, consistent with previously reported values~\cite{gong2024isotope, clua2023isotopic}. 

Having characterized $A_{zz}$, we proceed to extract the transverse hyperfine components, $\{ A^{(i)}_{xx}, A^{(i)}_{yy}, A^{(i)}_{xy} \}$, for all three nuclear spins. In contrast to the axial case, the leading-order effects of the transverse components are suppressed in the secular regime, where the $z$-axis ZFS is much larger than the transverse hyperfine strengths, and their inherently second-order nature makes them challenging to resolve~\cite{gao2022nuclear, gong2024isotope}. 

To overcome this issue, we {\it tilt} the quantization axis of the central sensor by applying an {\it in-plane} external field, $B_\perp$, which is perpendicular to and competes with the $z$-axis ZFS. In this rotated basis, the transverse hyperfine components acquire significant projections along the new quantization axis, thereby enhancing their measurable effects. As shown in Fig.~\ref{fig:Fig2}c, increasing $|B_\perp|$ transforms the hyperfine spectrum from a simple four-peak structure into broad, widely distributed resonances. Moreover, rotation of the vector magnetic field around the in-plane direction ($\theta = 90^\circ$) produces highly angle-sensitive ODMR variations, arising from a complex mixing of axial and transverse components (Fig.~\ref{fig:Fig2}d). Although this series of ODMR spectra could, in principle, serve as ``fingerprints'' for extracting the transverse components, the experimental spectra lack sufficient resolution due to power broadening, spectral diffusion from coupling to more distant spin baths, and other extrinsic noise sources.

To address this, we instead turn to the time-domain dynamics of a central spin. Here, we utilize the fact that hyperfine interactions cause decoherence of the quantum sensor over time by entangling it with the surrounding nuclear spin bath: tracing out the bath transforms a globally pure state into a locally mixed state. When hyperfine-induced entanglement involves only a few nuclear spins, the central electronic spin exhibits periodic coherence modulations~\cite{childress2006coherent, qiu2021nuclear}, driven by quantum interference between distinct energy scales set by different hyperfine couplings (Fig.~\ref{fig:Fig1}b). This provides an alternative route to access all components of the hyperfine interaction Hamiltonian.

Specifically, we employ a spin-echo sequence to isolate hyperfine-induced dephasing from bath-induced decoherence arising from inhomogeneous broadening. We work in a rotated basis by applying an in-plane field of $B_\perp = 20$~mT, which yields the engineered eigenstates, $\{ \ket{E_1} \approx \ket{0}, \ket{E_2} = \ket{M_-}, \ket{E_3} \approx \ket{M_+} \}$, where $\ket{M_\pm} = (\ket{+1} \pm e^{2i\phi} \ket{-1})/\sqrt{2}$ and $\phi$ denotes the azimuthal angle of $B_\perp$ relative to the crystal coordinates. We then define an effective two-level system spanned by $\{ \ket{\downarrow} = \ket{E_1}, \ket{\uparrow} = \ket{E_3} \}$ and measure the coherence, $C(T) = \langle \psi(T)|\hat{\sigma}_x|\psi(T)\rangle$, as a function of the interrogation time $T$, starting from an initially $x$-polarized state $\ket{\psi(0)} = (\ket{\downarrow} + \ket{\uparrow})/\sqrt{2}$ (Methods). Here, $\hat{\sigma}_x$ denotes the Pauli $x$ operator.

As shown in Fig.~\ref{fig:Fig2}e, we observe modulated coherence dynamics (\textcircled{\tiny I} and \textcircled{\tiny II}), accompanied by an overall envelope decay (\textcircled{\tiny III}). Physically, these three regimes correspond to entanglement with neighboring nuclear spins (\textcircled{\tiny I}), their subsequent disentanglement (\textcircled{\tiny II}), and complete decoherence (\textcircled{\tiny III}) as the sensor progressively couples to more distant spins in a larger bath and to extrinsic classical noise sources (Fig.~\ref{fig:Fig2}f).Formally, the decoherence dynamics, $C(T)$, can be analytically modeled as~\cite{childress2006coherent, qiu2021nuclear}
\begin{align}
    C(T) = \left[(1-c) + c \prod_{i=1}^3 Q_i(T) \right] e^{-(T/T_\text{1/e})^\beta}, \label{eq:ESEEM}
\end{align}
where $c$ represents the phenomenological modulation contrast, the exponential factor describes the envelope decay with $1/e$ time constant, $T_\text{1/e}$ and stretch exponent, $\beta$, and $Q_i(T)$ characterizes the periodic entanglement-disentanglement dynamics between the $i$-th nuclear spin and the central sensor. Crucially, $Q_i(T)$ depends on the full hyperfine tensor, $A^{(i)}_{\mu\nu}$, the magnitude of the in-plane magnetic field, $|B_\perp|$, and its orientation angles, $\theta$ and $\phi$, allowing full reconstruction of the hyperfine Hamiltonian through optimal numerical fitting (Methods).

Experimentally, we observe that the $C(T)$ modulation profiles are highly sensitive to the angle $\theta$ near $90^{\circ}$ (Fig.~\ref{fig:Fig2}g, left). Leveraging this pronounced sensitivity, we fit $C(T)$ to experimental data across all rotation angles, $\theta$, from which the hyperfine parameters shown in Fig.~\ref{fig:Fig2}h are extracted (Extended Data Fig.~\ref{fig:hyperfine}). We find that the decoherence dynamics predicted from the optimal hyperfine parameters show excellent agreement with both the representative line cut (black line, Fig.~\ref{fig:Fig2}e) and the full angular dependence of the data (Fig.~\ref{fig:Fig2}g, right). 

As a benchmark, fits using previously reported literature values~\cite{gong2024isotope, ivady2020ab} show systematically worse agreement with $C(T)$ compared to our extracted parameters (Fig.~\ref{fig:Fig2}i, Extended Data Fig.~\ref{fig:compare}). Interestingly, we find that the in-plane gyromagnetic ratio of the boron vacancy is $\gamma_\perp/2\pi \approx 19.6$~GHz/T, about 30\% smaller than its out-of-plane counterpart, $\gamma_z$, revealing anisotropic magnetic sensitivity in 2D hBN (black dashed line in Fig.~\ref{fig:Fig2}c). Consistent with this anisotropy, the ratio of the transverse hyperfine components between the DFT prediction~\cite{ivady2020ab} and our extracted value closely matches the gyromagnetic anisotropy here, suggesting an incorrect assumption underlying the {\it ab initio} computation.

The fully reconstructed hyperfine interaction Hamiltonian, obtained from short-time decoherence dynamics, provides a basis for developing quantum control protocols of neighboring nuclear spins, enabling their use as auxiliary quantum resources~\cite{marcks2025nuclear}. The reliability of our hyperfine Hamiltonian learning procedure is further validated using synthetic datasets, all of which converge to the ground-truth values (Supplementary Information).

\begin{figure*}[htbp!] 
\begin{center}
\includegraphics[scale=1]{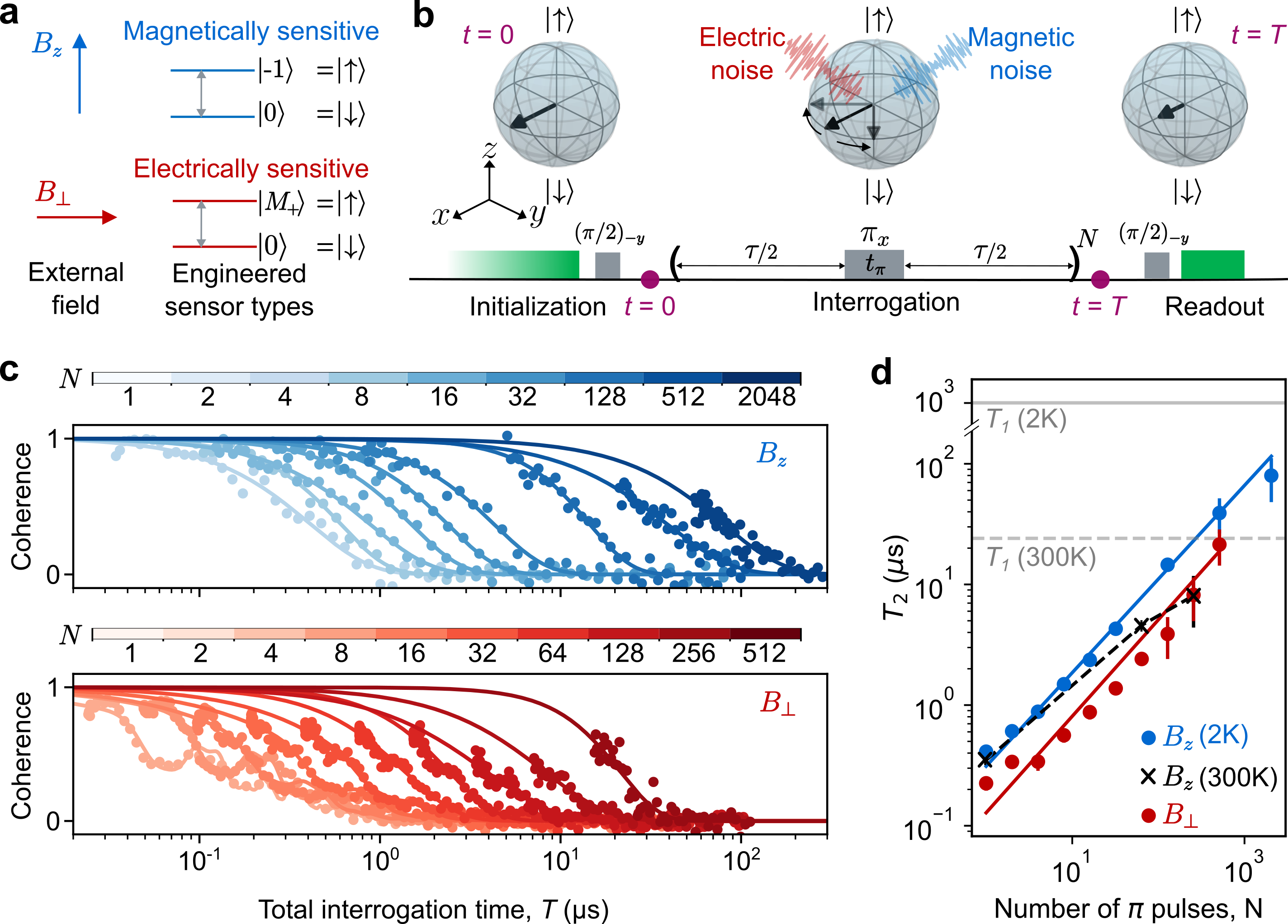}
\caption{\textbf{Probing magnetic and electric noise contributions to sensor decoherence.} 
\textbf{a,} The central spin switches between magnetic and electric noise sensing modes depending on the external field orientation: a field applied along the out-of-plane direction ($B_z$) realizes a magnetic-noise sensor with qubit states $\{\ket{-1}, \ket{0}\}$, whereas an in-plane field ($B_\perp$) realizes an electric-noise sensor with qubit states $\{\ket{M_+}, \ket{0}\}$. \textbf{b,} Sensor decoherence under magnetic and electric noise is independently probed using the CPMG sequence~\cite{meiboom1958modified, carr1954effects}, which applies $N$ spin-echo cycles over a total interrogation time $T = N(\tau + t_\pi)$, where $\tau$ is the variable spacing between adjacent $\pi$ pulses of duration $t_\pi$. An initial superposition along the $x$-axis of the Bloch sphere gradually loses coherence during $T$; this decoherence is then read out by mapping it into spin-state-dependent photoluminescence. \textbf{c,} CPMG decoherence profiles with varying number of $\pi$ pulses, $N$, are measured under magnetic (top) and electric (bottom) sensing modes. Solid lines denote stretched-exponential fits to each sequence, with hyperfine interaction-induced envelope modulation included in the $B_\perp$ case (Methods). \textbf{d,} At low temperature ($\approx2$K), the extracted $1/e$ coherence times, $T_2$, exhibit power-law scaling with $N$, following $T_2 \propto N^{0.78 \pm 0.03}$ and $T_2 \propto N^{0.72 \pm 0.04}$ for the $B_z$ and $B_\perp$ cases, respectively. At room temperature ($\approx 300$K), $T_2$ initially increases with $N$ but ultimately saturates due to the short $T_1$ limit (dashed line).}
\label{fig:Fig3}
\end{center}
\end{figure*}

\subsection*{Switchable Magnetic and Electric Noise Sensing}

To better understand the late-time decoherence seen in the spin-echo measurement (Fig.~\ref{fig:Fig2}e, \textcircled{\tiny III})—which ultimately limits the sensitivity of our central quantum sensor—we focus on identifying the underlying noise sources. In addition, establishing whether the observed decoherence arises from intrinsic (spin relaxation) or extrinsic (environmental noise) mechanisms is essential for assessing the prospects for room-temperature operation. In particular, to systematically separate magnetic and electric noise contributions, we exploit the ability to engineer the susceptibilities of the sensor’s eigenstates to different types of noise using an external field.

Concretely, when the field is aligned parallel to the ZFS, the isolated two-level states, $\{ \ket{\downarrow} = \ket{0}, \ket{\uparrow} = \ket{-1} \}$, can be selected to form a qubit whose transition frequency is sensitive to magnetic noise fluctuations, thereby realizing a magnetic-field sensor (top, Fig.~\ref{fig:Fig3}a). In contrast, applying the field orthogonal to the ZFS produces magnetically {\it insensitive} eigenstates, $\{ \ket{0}, \ket{M_+}, \ket{M_-} \}$, with vanishing magnetic moments to leading order (Methods); in this case, a qubit defined by $\{ \ket{\downarrow} = \ket{0}, \ket{\uparrow} = \ket{M_+} \}$ becomes sensitive to electric noise fluctuations, effectively acting as an electric-field sensor~\cite{dolde2011electric, huxter2023imaging} (bottom, Fig.~\ref{fig:Fig3}a).

To quantify the relative contributions of magnetic and electric noise to sensor decoherence, we perform dynamical decoupling measurements with the CPMG sequence~\cite{carr1954effects, meiboom1958modified} in both sensing modes (Fig.~\ref{fig:Fig3}b). The CPMG sequence consists of repeated spin-echo cycles within a fixed interrogation time $T$, with a variable number of $\pi$ pulses, $N$, which progressively suppresses slower noise components as $N$ increases~\cite{rizzato2023extending}. By comparing how the $1/e$ coherence times, $T_2$, scale with $N$ across the two sensing modalities, we quantitatively resolve how electromagnetic fluctuations over different spectral ranges contribute to sensor decoherence.

As shown in Fig.~\ref{fig:Fig3}c, increasing $N$ extends the coherence time for both noise types, exhibiting power-law scaling with $N$ (Fig.~\ref{fig:Fig3}d). Notably, in the magnetic-sensing case, we obtain $T_2 \approx 80~\mu$s for $N = 2048$ pulses—the longest electron-spin coherence time measured in any van der Waals material. Given a depolarization time of $T_1 \sim 1$ ms at low temperature~\cite{gong2024isotope}, the coherence time could, in principle, be further extended through more effective noise decoupling using a larger number of pulses. In contrast, at room temperature, $T_2$ is limited by a much shorter depolarization time of $T_1 \sim 10~\mu$s (cross markers, Fig.~\ref{fig:Fig3}d), underscoring the need to extend $T_1$ for room-temperature quantum applications.

\begin{figure*}[htbp] 
\begin{center}
\includegraphics[scale=1]{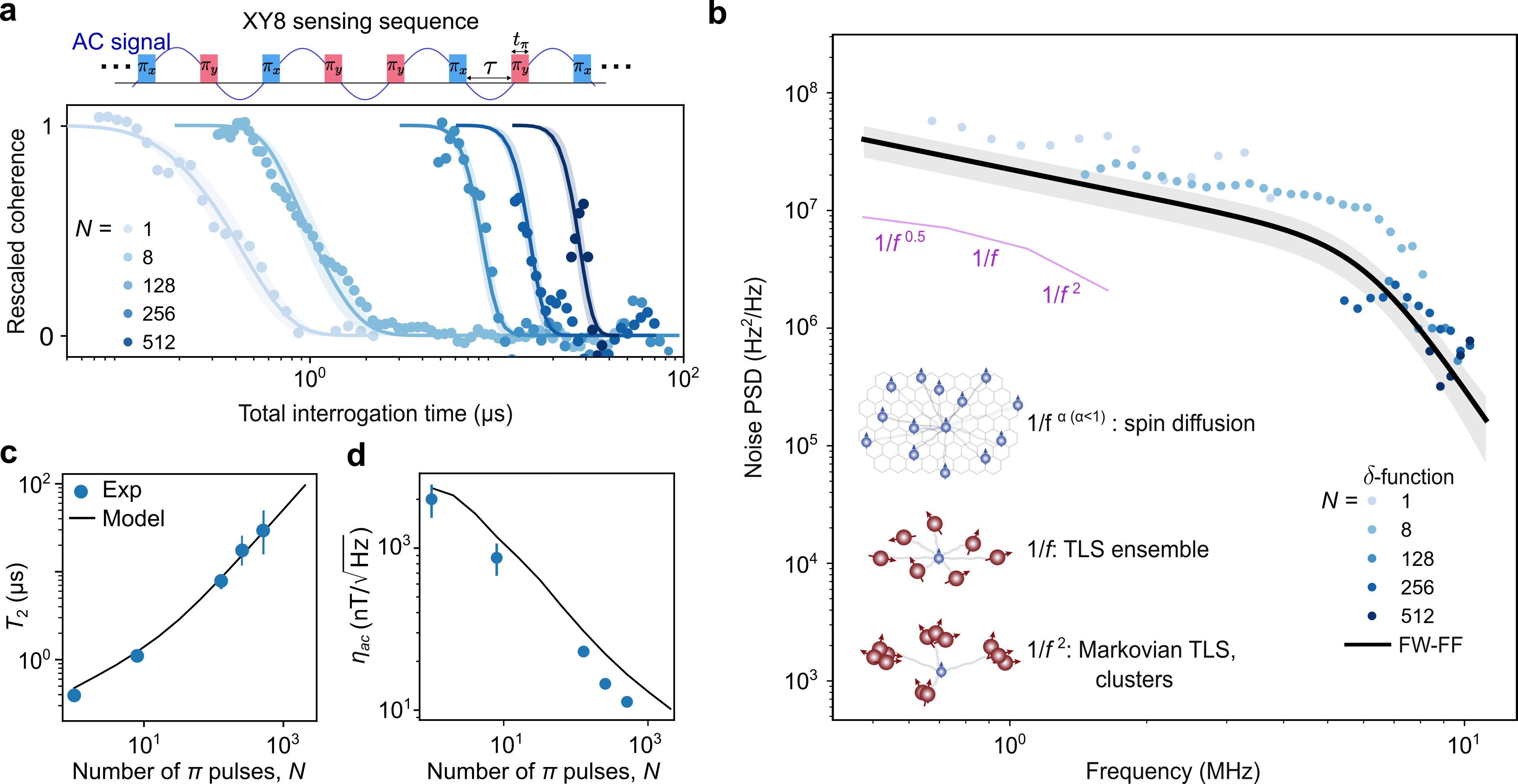}
\caption{\textbf{Noise spectrum reconstruction for an ensemble quantum sensor.} \textbf{a,} The XY8 sequence serves as an effective sensing sequence for detecting an AC signal with an oscillation period $T_\text{ac} = 2(\tau+t_\pi)$ (top). The phases of $\pi$ pulses in the XY8 sequence are chosen to alternate between the $x$- and $y$-axes for robustness against control pulse imperfections (Methods). Coherence profiles measured under XY8 in the magnetic sensing mode with $N = 1$, $8$, $128$, $256$, and $512$ pulses are shown, with appropriate contrast rescaling required by the filter-function formalism (see Extended Data Fig.~\ref{fig:sensitivity} for unscaled data). Solid lines represent predictions from the reconstructed noise PSD in \textbf{b}. \textbf{b,} Reconstructed noise PSD using a finite-width filter function (FW-FF) (black line), along with that reconstructed individually using an approximate filter function based on the $\delta$-function approximation (markers). Gray shading indicates the uncertainty of the noise PSD. See Methods for a detailed noise reconstruction procedure. Insets show various theoretical decoherence models predicting power-law scaling of noise spectra with different exponents (Methods). \textbf{c, d,} The model predictions based on the reconstructed noise PSD (solid lines) show consistent results for both coherence times (\textbf{c}), and estimated ac sensitivities (\textbf{d}) across all values of $N$.}
\label{fig:Fig4}
\end{center}
\end{figure*}

\subsection*{Noise Spectrum Reconstruction}

Periodic $\pi$-pulse trains used for coherence extension can also act as an effective sensing sequence for detecting AC fields when the field's oscillation period matches the sequence periodicity~\cite{degen2017quantum}. While a sensing sequence applied to ensemble-based sensors can boost sensitivity through parallel statistical averaging in space, it is often limited by disorder in the sensors’ transition energies, known as inhomogeneous broadening, which leads to control pulse imperfections and, consequently, additional decoherence. To mitigate this, we employ the XY8 sequence (top, Fig.~\ref{fig:Fig4}a)~\cite{gullion1990new}, which is robust against control errors, and numerically confirm that it achieves near-ideal sensing performance in disordered spin ensembles compared to CPMG (Extended Data Fig.~\ref{fig:robustness}).

Using the XY8 sensing sequence, we now turn to characterizing the magnetic noise bath of our 2D spin ensemble sensor. Specifically, under XY8 with $N$ $\pi$ pulses, the coherence at interrogation time $T$, $C_N(T)$, is related to the noise PSD, $S(\omega)$, as
\begin{align}
    -\log C_N(T) = \frac{1}{\pi} \int _{0}^{\infty} d\omega S(\omega) \frac{F_N(\omega, T)}{\omega^2}. \label{eq:chi(T)}
\end{align}
Here, $F_N(\omega, T)$ is the realistic, finite-width filter function (FW-FF) that explicitly accounts for finite pulse duration effects, ensuring accuracy in the high control duty-cycle regime~\cite{ishikawa2018influence} (Extended Data Fig.~\ref{fig:filter}). Since $F_N(\omega, T)$ is analytically known and $C_N(T)$ is obtained from experimental measurements, the noise PSD can be reconstructed by solving the inversion problem using a numerical optimization algorithm (Methods).

Figure~\ref{fig:Fig4}a shows coherence profiles measured under XY8 sequences with varying numbers of $\pi$ pulses. Each profile starts at $T = N t_\pi$, the earliest time point where free evolution occurs between adjacent $\pi$ pulses, and is rescaled to unity at that point, consistent with the filter function assumption that decoherence during individual pulses is neglected (see Extended Data Fig.~\ref{fig:sensitivity}a for unscaled coherence). 

Using these datasets, together with the corresponding FW-FFs, we reconstruct the noise PSD, shown as the black line in Fig.~\ref{fig:Fig4}b (see Methods for a detailed description of the reconstruction procedure based on individual XY8 time traces). To highlight the importance of using an accurate filter function, we compare the reconstructed noise PSD with that obtained using an approximate filter function, in which $F_N(\omega,T)$ is modeled as a delta function peaked at the sequence resonance frequency (data markers, Fig.~\ref{fig:Fig4}b). Although the delta-function approximation qualitatively captures the overall shape of the noise PSD, it results in a quantitative amplitude mismatch due to the neglect of finite pulse-width effects and higher-order resonance peaks (Methods).

The reconstructed noise PSD for our ensemble sensor follows a power-law scaling of $1/\omega^\alpha$ with $\alpha \approx 0.9$, before rolling off more steeply near $\sim$10~MHz (black line, Fig.~\ref{fig:Fig4}b). Various theoretical decoherence models for a central spin system, such as spin diffusion, coupling to an ensemble of two-level systems (TLS), or Markovian clusters of TLS, predict power-law decays with exponents in the range $0 < \alpha \le 2$ (inset, Fig.~\ref{fig:Fig4}b; see Methods for a detailed discussion of these possible noise sources).

We confirm that the reconstructed noise PSD not only captures the $C_N(T)$ profiles across all XY8 sequences (solid lines, Fig.~\ref{fig:Fig4}a), but also quantitatively reproduces the measured $T_2$ values (Fig.~\ref{fig:Fig4}c) and estimated ac sensitivities $\eta_{\text{ac}}$ (Fig.~\ref{fig:Fig4}d), demonstrating the consistency of our noise model with experimental observations. We further validate our noise-learning results with full quantum-mechanical simulations, which yield consistent outcomes (Extended Data Fig.~\ref{fig:decoupling}). 

The electric-field noise PSD could, in principle, be reconstructed similarly by deconvolving the coherence envelope modulation arising from hyperfine interactions with the nuclear spin bath to isolate pure electric-field noise. However, this requires complete knowledge of the hyperfine coupling strengths of more distant nuclear spins, which also interact slowly and coherently with the sensor spins. Incorporating these contributions provides a natural path to a more accurate reconstruction of the electric-field noise PSD.

\subsection*{Discussion on Sensitivity and Outlook}

A key benchmark for characterizing quantum sensing performance is the best achievable sensitivity~\cite{barry2020sensitivity}. From our measurements using an ensemble of $V_{B}^{-}$ defects in 2D hBN, we estimate an AC magnetic sensitivity of $\eta_\text{ac} \approx 138~\mathrm{nT}/\sqrt{\mathrm{Hz}}$ at $T\approx 2$K and $\eta_\text{ac} \approx 290~\mathrm{nT}/\sqrt{\mathrm{Hz}}$ at room temperature (Methods, Fig.~\ref{fig:Fig1}c). This represents the best performance reported to date among van der Waals materials and places our ensemble sensor on par with leading systems such as NV center ensembles in diamond.

Assuming that an ensemble sensor detects a globally correlated signal, its sensitivity scales as $\eta_\text{ac} \propto 1/\mathcal{C}\sqrt{N_e T_2}$, where $\mathcal{C}$ denotes the single-spin detection efficiency and $N_e$ is the number of non-interacting quantum sensors~\cite{barry2020sensitivity}. This suggests several clear routes for further improvement. First, our present detection scheme does not fully exploit the advantages of large-$N_e$ ensembles due to the saturation of a single-photon counting module (Methods). We anticipate that an improved detector capable of fully utilizing the ensemble signal would significantly enhance the sensitivity, potentially reaching the nanotesla level (Extended Data Figs.~\ref{fig:sensitivity}b,c). Second, extending $T_2$ requires suppressing residual electromagnetic noise, even though surface effects are naturally mitigated in atomically thin 2D materials. Promising approaches include post-fabrication cleaning~\cite{rosenberger2018nano}, embedding defect layers within pristine 2D heterostructures~\cite{wang2023clean, purdie2018cleaning}, and active stabilization of spin-resonance transitions~\cite{akbari2022lifetime}. Finally, tailoring the emission profile of these atomic-scale single-photon emitters via integration with on-chip photonics~\cite{li2021integration, gelly2023inverse} could dramatically improve collection efficiency while simultaneously providing a scalable route to device integration.

Taken together, our device-agnostic Hamiltonian- and noise-learning framework will accelerate the discovery of novel quantum sensing platforms based on atomically thin 2D van der Waals materials, as well as other central spin systems. By integrating nanoscale sensing with coherent, tunable hybrid electron-nuclear registers, these 2D spin systems open pathways not only to advanced quantum sensing but also to scalable quantum simulation~\cite{rovny2024nanoscale} and quantum networking~\cite{reiserer2022colloquium}, heralding a new era of fully integrated solid-state quantum technologies.
\\

\textit{Acknowledgments}--- We thank Sungjun Eun for assistance with running simulations on Marlowe, Stanford’s GPU-based computational cluster; Timothy Chang, Indra Periwal and Jakob Grzesik for support with initial optical and spin measurements. We also thank Norman Yao for helpful discussions. This work was supported by the Department of Energy DE-SC0025620, Laboratory Directed Research and Development program at SLAC National Accelerator Laboratory, under contract DE-AC02-76SF00515, and the Quantum Science Seed Grant Program from Stanford Q-FARM. JC was supported by AFOSR under grant no. FA9550-23-1-0625 and AFOSR YIP under grant no. FA9550-25-1-0147. We also acknowledge support from the Stanford Research Computing Center for providing computational resources, as some of the computations for this project were performed on the Sherlock cluster. G.S. acknowledges support from the Stanford Bloch Postdoctoral Fellowship. E.I.R. acknowledges support from an appointment to the Intelligence Community Postdoctoral Research Fellowship Program at Stanford University, administered by the Oak Ridge Institute for Science and Education (ORISE) through an interagency agreement between the U.S. Department of Energy and the Office of the Director of National Intelligence (ODNI). Support for hBN crystal growth by TP and JHE was provided by the Office of Naval Research, award number N00014-22-1-2582. J.C. acknowledges support from the Terman Faculty Fellowship at Stanford University.

\clearpage
\section*{Methods}
\subsection{Experimental setup}

Extended Data Fig.~\ref{fig:setup} illustrates the experimental setup used in this work. The device under test is measured in a closed-cycle cryostat (attoDRY2100, Attocube) equipped with a two-axis vector magnet and a base temperature of $\approx 2\mathrm{K}$. Optical excitation and collection are carried out in free space using a home-built confocal microscope, which includes a light-emitting diode (MF617, Thorlabs) and a camera for sample imaging. A cryogenic objective (LT-APO/VISIR/0.82, Attocube) focuses the light onto the sample. Sample scanning in all directions is achieved with Attocube nanopositioners (ANPx101 for lateral positioning and ANPz101 for vertical positioning).

Off-resonant spin-state initialization is performed with a 532~nm laser (split from a Verdi V-10, Coherent), gated by an acousto-optic modulator (4C2C-532-AOM, Gooch and Housego). A 550~nm short-pass filter in the excitation path suppresses low-frequency, fiber-induced photoluminescence and Raman emission. The collected signal is filtered through a 633~nm long-pass filter, a 532~nm notch filter, and a 900~nm short-pass filter to isolate the hBN phonon sideband. The excitation and collection paths are separated by a dichroic mirror (Di02-R561-25x36, Semrock), which provides additional short-pass filtering for excitation and long-pass filtering for collection. Finally, the fluorescence from the free-space optical setup is coupled into a single-mode fiber (780HP, Thorlabs) and directed to a single-photon counting module.

Microwave control is implemented using a vector signal generator (SG396, Stanford Research Systems), driven by arbitrary waveform generators (Pulse Streamer 8/2, Swabian, and HDAWG, Zurich Instruments). Single-photon detection is performed using a single-photon counting module (SPCM-AQRH-WX-TR, Excelitas). Microwave switches (ZASWA-2-50DRA+, Mini-Circuits) with a rise time of 20~ns are used to gate both the photon-signal acquisition and the microwave output (only for ODMR experiments). Pulsed experiments employ direct IQ modulation on the signal generator to improve pulse sharpness. Microwave signals are amplified before entering the cryostat by a high-power (50 W) amplifier (50S1G4AM2, Amplifier Research).

Inside the cryostat, 0 dB cryogenic attenuators (2082-6418-dB-CRYO, XMA) are installed on both the input and output microwave lines to thermalize the center conductors through improved metal-to-metal contact with the outer conductor. The output line is terminated with a high-power 50-Ohm load. Pulse sequences are generated with the \texttt{nspyre} framework (University of Chicago) and the \texttt{zhinst-toolkit} API (Zurich Instruments). Microwave lines are wire-bonded to a printed circuit board (PCB) through surface-mount launchers that connect to bonding pads. The configuration used in this work builds on a previous generation of the setup~\cite{rosenthal:2023, rosenthal:2024}.

\subsection{Device fabrication}

For details on the growth of isotopically purified hBN, see Ref.~\cite{gong2024isotope}. hBN flakes are exfoliated from bulk crystals using the standard scotch-tape method and transferred onto a SiO$_2$/Si substrate. The substrate is pretreated with O$_2$ plasma (50~W, 20~sccm) for 1~min to enhance adhesion of flakes during exfoliation. Adhesion is further improved by heating the substrate to $100^{\circ}\mathrm{C}$ during exfoliation, after which the tape is peeled off.

$^4$He$^+$ ion implantation is performed at CuttingEdge Ions LLC using a 3~keV beam energy and a dose of 1~ion/nm$^2$ to generate boron-vacancy $V_B^-$ defects. This nominally corresponds to a defect density of $\sim$150~ppm, where interactions between spin defects may not be negligible. However, our coherence measurements using the XY8 sequence show a continuously increasing $T_2$, exceeding $\approx$30~$\mu$s. This establishes a lower bound on the sensor-sensor distance that exceeds the total sample thickness; if the coherence were limited by dipolar interactions between boron-vacancy defects, the $T_2$ scaling would instead be expected to saturate~\cite{zhou2020quantum}. This supports the effectively 2D nature of the sample and suggests that the actual defect density is significantly lower, likely due to the low yield of $V_B^-$ generation. The exact origin of this density mismatch requires further investigation, which we leave for future work.

Following implantation, suitable hBN flakes (5–30~nm thick) are transferred onto a coplanar microwave waveguide chip using a standard polycarbonate-assisted transfer method. The data presented in this paper correspond to measurements performed on a 10 nm thick sample, as confirmed by atomic force microscopy. The 10-nm-thick sample chip is then mounted on a custom-designed PCB and wire-bonded. For the fabrication of the coplanar waveguide, a sapphire chip is diced into 5~mm~$\times$~5~mm squares. A 400~nm-thick layer of electron-beam resist (PMMA950A4) is spin-coated, and the pattern is defined using electron-beam lithography (Raith EBPG5200). Resist development is carried out in MIBK:IPA (1:3) for 1 minute, followed by rinsing in isopropyl alcohol (IPA). Metal deposition of 10~nm Ti and 300~nm Au is then performed, followed by liftoff in Remover PG, acetone, and IPA.

\subsection{ODMR spectroscopy}
We employ ODMR spectroscopy to probe the energy spectrum in the GS manifold of the boron-vacancy defect. The ODMR protocol consists of continuous-wave optical excitation with an off-resonant 532 nm green laser, combined with microwave driving at a variable carrier frequency swept across successive experiments. Specifically, we begin by applying a $10~\mu\mathrm{s}$ green laser pulse to polarize the defect’s electron spin into the $\ket{0}$ state, thereby establishing a baseline for the photoluminescence signal from the central spin (in other control sequence measurements the initialization period is shortened to 0.4~$\mu$s). We then repeat the same optical pulse, this time accompanied by a simultaneous microwave pulse. When the microwave carrier frequency is resonant with a spin transition, the spin population is redistributed among the GS sublevels, leading to a reduction in photoluminescence. The difference in photon counts between the reference and microwave-driven sequences defines the ODMR contrast, which is typically on the order of 10\% and reveals the energy eigenstructure of the ground-state manifold.

\subsection{Hamiltonian of the 2D hBN spin system}

In our experiments, we probe an ensemble of boron-vacancy defects that collectively serve as a quantum sensor. Due to the crystalline structure of the hBN lattice, each boron-vacancy defect is embedded in a {\it deterministic} nuclear spin environment, forming a central spin system. Specifically, the Hamiltonian of this central spin system can be written as $\hat{H} = \hat{H}_0 + \hat{H}_\text{HF}$, where $\hat{H}_0$ denotes the dominant electronic spin Hamiltonian of a boron-vacancy defect,
\begin{align}
\hat{H}_0 &= D \hat{S}_z^2 + \gamma_z B_z \hat{S}_z + \gamma_\perp (B_x \hat{S}_x + B_y \hat{S}_y) \nonumber \\
&\quad + \mathcal{E}_1 (\hat{S}_x^2 - \hat{S}_y^2) + \mathcal{E}_2 (\hat{S}_x \hat{S}_y + \hat{S}_y \hat{S}_x), \label{eq:H0}
\end{align}
with $\hbar = 1$. Here, $D/2\pi \approx 3.65$~GHz is the ZFS, $\gamma_z/2\pi \approx 28$~GHz/T and $\gamma_\perp/2\pi \approx 19.6$~GHz/T are the out-of-plane and in-plane gyromagnetic ratios of the boron vacancy, respectively, $\mathcal{E}_{1,2}$ are the local crystal strain strengths, which also include contributions from electric fields, and $\hat{S}_{x,y,z}$ are the spin-1 operators for the three magnetic sublevels of the GS manifold. The components of the external magnetic field, $\vec{B} = (B_x, B_y, B_z)$, are given by $B_x = |\vec{B}| \sin\theta \cos\phi$, $B_y = |\vec{B}| \sin\theta \sin\phi$, and $B_z = |\vec{B}| \cos\theta$, and can be tuned by varying the orientation angles, $\phi$ and $\theta$, relative to the hBN lattice, as well as by adjusting the amplitude, $|\vec{B}|$, of the applied field.

$\hat{H}_\text{HF}$ represents the perturbative hyperfine interaction with the three nearest-neighbor $^{15}$N nuclear spins, as given in Eq.~(\ref{eq:H_HF}) of the main text. $\hat{H}_\text{HF}$ is an approximate Hamiltonian that neglects longer-range but weaker interactions with the more distant spin bath, since their large energy mismatch causes their contribution to manifest primarily as inhomogeneous broadening of the central spin sensor. The nuclear Zeeman splitting and nuclear dipole–dipole interactions of the bath are also neglected, as their timescales are too short to affect the experimental results reported here. In Sec.~\ref{hahn:analysis}, however, we include the nuclear Zeeman splitting for completeness.

\subsection{Analysis of zero-field sensor eigenstates}

The zero-field ODMR spectrum enables the extraction of both the local crystal strain strength and the axial hyperfine component. Specifically, under the secular approximation, the zero-field Hamiltonian is given as $\hat{H} = D \hat{S}_z^2 + \mathcal{E}_1 (\hat{S}_x^2 - \hat{S}_y^2) + \mathcal{E}_2 (\hat{S}_x \hat{S}_y + \hat{S}_y \hat{S}_x) + A_{zz} \hat{S}_z \hat{I}_{z,\text{tot}}$, with $\hat{I}_{z,\text{tot}} = \sum_{i=1}^3 \hat{I}_z^{(i)}$. Since $\hat{I}_{z,\text{tot}}$ can be treated as a discrete random variable taking the four values of $m_{z,\text{tot}} = -\frac{3}{2}, -\frac{1}{2}, \frac{1}{2}, \frac{3}{2}$, the Hamiltonian can be rewritten as $\hat{H}(m_{z,\text{tot}}) = D \hat{S}_z^2 + \mathcal{E}_1 (\hat{S}_x^2 - \hat{S}_y^2) + \mathcal{E}_2 (\hat{S}_x \hat{S}_y + \hat{S}_y \hat{S}_x) + A_{zz} \hat{S}_z m_{z,\text{tot}}$, which can be represented as a reduced $3 \times 3$ matrix in the spin-1 basis. Diagonalizing $\hat{H}(m_{z,\text{tot}})$ yields the $m_{z,\text{tot}}$-dependent energy eigenvalues, $E_1 = 0$, $E_2 = D - \sqrt{(A_{zz} m_{z,\text{tot}})^2 + \mathcal{E}^2}$, and $E_3 = D + \sqrt{(A_{zz} m_{z,\text{tot}})^2 + \mathcal{E}^2}$, where $\mathcal{E} = \sqrt{\mathcal{E}_1^2 + \mathcal{E}_2^2}$ denotes the combined strain strength. The corresponding resonance transition frequencies are then given by
\begin{align}
    f_{\pm} (m_{z,\text{tot}}) &= (D \pm \sqrt{(A_{zz} m_{z,\text{tot}})^2 + \mathcal{E}^2})/2\pi,
\end{align}
which produces a four-peak ODMR spectrum that is doubly degenerate, i.e., independent of the sign of $m_{z,\text{tot}}$. Note that although the zero-field ODMR spectrum appears qualitatively similar to that measured under a finite $B_z$ field, with both exhibiting four hyperfine peaks, the spacing between the resonance peaks is modified by the strain contribution. Using the value of $A_{zz}$ obtained from ODMR measurements at non-zero $B_z$, we fit the zero-field ODMR spectrum to obtain the strain term $\mathcal{E}/2\pi = 18.5 \pm 0.4~\text{MHz}$, and zero-field splitting $D/2\pi = 3.653 \pm0.016$~GHz (Supplementary Information).

\subsection{Analysis of engineered sensor eigenstates}

Here, we provide a detailed analysis of how the spin-1 GS manifold is manipulated by the external vector magnetic field and thus responds differently to hyperfine interactions, as well as how dynamical decoupling under different field orientations enables switchable sensing between magnetic and electric noise.

For the case where the applied field is aligned with the $z$-axis, parallel to the out-of-plane axis of the hBN crystal ($\theta = 0^\circ$), the dominant electronic spin Hamiltonian can be approximated as $\hat{H}_0 \approx D \hat{S}_z^2 + \gamma_z B_z \hat{S}_z$, since the large $D$ along the $z$-axis defines the quantization axis. Consequently, the eigenstates of the spin-1 $\hat{S}_z$ operator remain eigenstates of $\hat{H}_0$: $\{\ket{E_1} = \ket{0}, \ket{E_2} = \ket{-1}, \ket{E_3} = \ket{+1}\}$, with corresponding energy eigenvalues $E_1 = 0$, $E_2 = D - \gamma_z B_z$, and $E_3 = D + \gamma_z B_z$. We consider two states, $\{\ket{\downarrow} = \ket{E_1}, \ket{\uparrow} = \ket{E_2}\}$, which serve as an effective qubit, with the transition frequency, $f$, given by $f = (D - \gamma_z B_z)/2\pi$. Under the hyperfine interactions within the secular approximation, the transition frequency splits into four distinct values, $f = (D - \gamma_z B_z + A_{zz} m_{z,\text{tot}})/2\pi$ with $m_{z,\text{tot}} = -\frac{3}{2}, -\frac{1}{2}, \frac{1}{2}$, and $\frac{3}{2}$. By measuring the spectral spacing between the hyperfine peaks at different $B_z$ values, we extract both the axial hyperfine component, $A_{zz}/2\pi = -67 \pm 0.5$~MHz, and the $z$-axis gyromagnetic ratio, $\gamma_z/2\pi = 28 \pm 0.2$~GHz/T (Supplementary Information).

Importantly, fluctuations in the transition frequency, $\delta f$, are governed by magnetic field noise projected along the $z$-axis, $\delta B_z$, such that $\delta f = \gamma_z \delta B_z / 2\pi$, which leads to decoherence of the central boron-vacancy qubit. Assuming no fluctuations in the total nuclear spin magnetization over the interrogation time, dephasing due to secular hyperfine interactions can be effectively suppressed by dynamical decoupling. This indicates that boron-vacancy defects under dynamical decoupling can be used to probe decoherence arising from external magnetic noise sources in this configuration (Fig.~\ref{fig:Fig3} of the main text).

In contrast, when the applied field is perpendicular to the $z$-axis ($\theta = 90^\circ$), and the in-plane field strength, $B_\perp$, is larger than the strain but smaller than the ZFS, i.e., $D \gg \gamma_\perp B_\perp \gg \mathcal{E}_{1,2}$, the electronic spin Hamiltonian can be approximated as $\hat{H}_0 \approx D \hat{S}_z^2 + \gamma_\perp (B_x \hat{S}_x + B_y \hat{S}_y)$, with $B_x = B_\perp \cos \phi$ and $B_y = B_\perp \sin \phi$. Note that our experimental condition ($B_\perp\sim20$~mT) satisfies this assumed energy hierarchy, with $D/2\pi \approx 3.65~\text{GHz} \gg \gamma_\perp B_\perp/2\pi \approx 0.39~\text{GHz} \gg \mathcal{E}_{1,2}/2\pi \sim 0.01~\text{GHz}$. The corresponding energy eigenstates and eigenvalues are then given by
\begin{align}
    \ket{E_1} &= \cos \alpha \ket{0} - \sin\alpha \ket{M_+}, \\
    \ket{E_2} &= \ket{M_-}, \\
    \ket{E_3} &= \cos \alpha \ket{M_+} + \sin \alpha \ket{0}, 
\end{align}
with $E_1 = \frac{1}{2}(D - \sqrt{4 \gamma_\perp^2 B_\perp^2 + D^2})$, $E_2 = D$, and $E_3 = \frac{1}{2}(D + \sqrt{4 \gamma_\perp^2 B_\perp^2 + D^2})$, respectively, and $\tan 2\alpha = \frac{2 \gamma_\perp B_\perp}{D}$. Note that when $D \gg \gamma_\perp B_\perp$, corresponding to $\cos \alpha \approx 1$ and $\sin \alpha \approx 0$, the eigenstates can be further approximated as $\{\ket{E_1} \approx \ket{0},\ \ket{E_2} = \ket{M_-},\ \ket{E_3} \approx \ket{M_+}\}$. Here, the two states, $\ket{M_\pm}$, are defined as
\begin{align}
\ket{M_\pm} &= \frac{\ket{+1} \pm e^{2i\phi}\ket{-1}}{\sqrt{2}},
\end{align}
and have zero magnetic moment, i.e., $\langle M_\pm | \hat{S}_z | M_\pm \rangle = 0$. Since $\langle 0 | \hat{S}_z | 0 \rangle = 0$ as well, all eigenstates carry no magnetic moment, rendering them insensitive to magnetic noise to leading order. 

With off-resonant green excitation predominantly populating $\ket{E_1} \approx \ket{0}$, the electronic spin transitions occur along two branches, $\ket{E_1} \leftrightarrow \ket{E_2}$ and $\ket{E_1} \leftrightarrow \ket{E_3}$, at transition energies $\Delta E_{21} \approx D + \frac{\gamma_\perp^2 B_\perp^2}{D}$ and $\Delta E_{31} \approx D + \frac{2\gamma_\perp^2 B_\perp^2}{D}$, consistent with the measured ODMR shown in Fig.~\ref{fig:Fig2} of the main text. To accurately extract the in-plane gyromagnetic ratio, $\gamma_\perp$, we diagonalize $\hat{H}_0$ including the strain term and fit the resulting resonances to the experimental data, yielding $\gamma_\perp/2\pi = 19.6 \pm 0.5~\text{GHz/T}$ (see the two black dashed lines in Fig.~\ref{fig:Fig2}c).

In this in-plane field configuration, we define the qubit states as $\{\ket{\downarrow} = \ket{E_1}, \ket{\uparrow} = \ket{E_3}\}$. As shown in Fig.~\ref{fig:Fig2}d, the electronic spin transitions are significantly broadened by hyperfine coupling to the three neighboring nuclear spins, producing numerous non-degenerate states with small splitting. Unlike the $\theta = 0^\circ$ case, the secular approximation is not valid at $\theta = 90^\circ$, and a simple $\pi$ pulse cannot dynamically decouple the qubit dephasing induced by the hyperfine Hamiltonian, leading to the strong modulation in the spin-echo profile. 

Beyond this rapid modulation of the coherence profile at short times---which also persists under dynamical decoupling sequences with a larger number of $\pi$ pulses---we observe slow dynamics that govern the decay of the coherence envelope at longer times. We attribute this late-time decoherence to time-dependent fluctuations of the qubit transition frequency arising from extrinsic noise. Specifically, when the $\mathcal{E}_{1,2}$ terms are explicitly retained in $\hat{H}_0$, they produce a small splitting between $\ket{E_2}$ and $\ket{E_3}$ proportional to $\mathcal{E} = \sqrt{\mathcal{E}_1^2 + \mathcal{E}_2^2}$, which, in turn, shifts the qubit transition frequency between $\ket{E_1}$ and $\ket{E_3}$~\cite{dolde2011electric}. Note that $\mathcal{E}$ includes contributions from both external electric noise and crystal strain~\cite{dolde2011electric}. Under the assumption of static strain, dephasing due to strain is dynamically decoupled, leaving time-dependent electric noise as the dominant source of decoherence. This indicates that the in-plane field configuration enables boron-vacancy defects to operate as effective electric field sensors, assuming that in-plane magnetic-field noise, $\delta B_\perp$, is not a leading-order contribution due to its suppression by a factor of $\frac{\gamma_\perp B_\perp}{D} \ll 1$.

\subsection{Rabi oscillation dynamics of the central spin}
Once the resonance transition frequencies are identified from the ODMR measurements, coherent Rabi oscillations can be driven between the GS spin eigenstates. As shown in Extended Data Fig.~\ref{fig:rabi}, we present the Rabi oscillation dynamics of the central boron-vacancy spin under two orientations of the external magnetic field: $B_{z}$ at $\theta = 0^\circ$ and $B_{\perp}$ at $\theta = 90^\circ$. 

The experimental Rabi oscillation signals, measured as a function of the microwave driving duration $t$, are normalized by the reference photon count without a microwave drive, yielding the Rabi signal contrast $c_\text{Rabi}(t)$. These data are then fitted to an exponentially decaying sinusoidal function,
\begin{align}
c_{\text{Rabi}}(t) &= -c_0 \left[\frac{1}{2} - \frac{1}{2} \cos(\Omega t) e^{-t/T_\text{Rabi}}\right],
\end{align}
where $c_0$ is the maximum contrast, $\Omega$ is the effective Rabi frequency, and $T_\text{Rabi}$ represents the $1/e$ decay time of the oscillations. Note that the negative contrast reflects a reduction in photon counts relative to the reference signal. The normalized contrast is then defined as $c_{\text{Rabi}}(t)/c_0$. 

For an out-of-plane $B_z$ field, where the hyperfine interaction produces a simple, well-resolved four-peak splitting, we find that the normalized Rabi oscillation decay time, $\Omega T_\text{Rabi}/2\pi$, is essentially independent of the Rabi frequency, with $\Omega T_\text{Rabi}/2\pi \approx 6$. This scale-invariant decay time indicates that the Rabi decay is predominantly governed by shot-to-shot fluctuations in the Rabi frequency. In contrast, for an in-plane $B_\perp$ field, where the hyperfine interaction gives rise to a broad distribution of complex resonance spectra, the Rabi decay time becomes highly sensitive to the choice of microwave carrier frequency at a given field strength. When the microwave drive excites multiple eigenstate transitions with different hyperfine-split detunings, the resulting Rabi oscillations exhibit significantly shorter lifetimes due to dephasing among eigenstates oscillating at different rates (Extended Data Fig.~\ref{fig:rabi}).

\subsection{Analysis of spin-echo envelope modulation}
\label{hahn:analysis}
Here, we describe the numerical simulation and fitting procedure used to analyze the coherence envelope modulation observed in the spin-echo measurements of Fig.~\ref{fig:Fig2} in the main text. Microscopically, the spin-echo envelope modulation arises from entanglement–disentanglement dynamics driven by coherent hyperfine interactions with neighboring nuclear spins, a phenomenon that has been investigated both theoretically and experimentally in various solid-state spin systems~\cite{childress2006coherent, qiu2021nuclear}. 

In particular, an analytical result~\cite{childress2006coherent} predicts that the central spin’s coherence modulation takes the form
\begin{align}
    C(T) = \prod_{i=1}^3 Q_i(T),
\end{align}
and that, after including phenomenological contrast reduction and decoherence effects, the full profile is given by Eq.~(\ref{eq:ESEEM}). Here, $Q_i(T)$ denotes the contribution from the $i$-th nuclear spin, given by
\begin{align}
    Q_i(T) &= 1 - 2 \left| {\vec\omega}^{(i)}_0 \times \vec\omega^{(i)}_+ \right|^2 \sin^2\left(\frac{\omega^{(i)}_0 T}{4}\right) \sin^2\left(\frac{\omega^{(i)}_+ T}{4}\right).
\end{align}
Here, $\vec{\omega}^{(i)}_\lambda$ and $\omega^{(i)}_\lambda$, with $\lambda = 0, -, +$, are the unit vectors along the quantization axis and the resonance transition frequencies, respectively, of the following Hamiltonian for the $i$-th nuclear spin:
\begin{align}
    \hat{H}_{\text{nuc},\lambda}^{(i)} = \gamma_n \vec{B} \boldsymbol{\cdot} \vec{\hat{I}}^{(i)} + \sum_{\mu,\nu =x,y,z} \langle \hat{S}_{\mu} \rangle_\lambda A^{(i)}_{\mu\nu} \hat{I}^{(i)}_{\nu}, \label{eq:H_nuc}
\end{align}
where $\gamma_n/2\pi = -4.3~\text{MHz/T}$ is the gyromagnetic ratio of the $^{15}$N nuclear spin, $\vec{B}$ is the three-dimensional vector magnetic field, and $\vec{\hat{I}}^{(i)} = (\hat{I}^{(i)}_x, \hat{I}^{(i)}_y, \hat{I}^{(i)}_z)$ is the vectorized spin-1/2 operator for the nuclear spin at site $i$. The expectation values of the electronic spin-1 operators are given by $\langle \hat{S}_{\mu} \rangle_0 = \langle E_1 | \hat{S}_\mu | E_1 \rangle$, $\langle \hat{S}_{\mu} \rangle_- = \langle E_2 | \hat{S}_\mu | E_2 \rangle$, and $\langle \hat{S}_{\mu} \rangle_+ = \langle E_3 | \hat{S}_\mu | E_3 \rangle$, evaluated using the energy eigenstates, $\ket{E_{1,2,3}}$, of the electronic spin Hamiltonian $\hat{H}_0$ given in Eq.~(\ref{eq:H0}), ordered such that $E_1 < E_2 < E_3$. 

Note that $\hat{H}_{\text{nuc},\lambda}^{(i)}$ includes the small nuclear Zeeman term for accuracy and can be represented as a $2 \times 2$ matrix for diagonalization. To gain an intuitive understanding, $\hat{H}_{\text{nuc},\lambda}^{(i)}$ can be rewritten as
\begin{align}
    \hat{H}_{\text{nuc},\lambda}^{(i)} &= \sum_{\nu=x,y,z} ({\Omega}_{\text{nuc}, \lambda} ^{(i)})_\nu \hat{I}^{(i)}_\nu,
\end{align}
where $({\Omega}_{\text{nuc}, \lambda} ^{(i)})_\nu$ is the $\lambda$-dependent nuclear Rabi frequency for the $i$-th spin along the $\nu$ direction, defined as $({\Omega}_{\text{nuc}, \lambda} ^{(i)})_\nu = \gamma_n B_\nu + \sum_{\mu=x,y,z} \langle \hat{S}_{\mu} \rangle_\lambda A^{(i)}_{\mu\nu}$. Essentially, the electronic spin polarization acts as a local magnetic field, inducing effective Rabi precession of the nuclear spin. In this picture, $\vec{\omega}^{(i)}_\lambda$ and $\omega^{(i)}_\lambda$ can be regarded as the Larmor precession axis and frequency, respectively. With the electronic spin initialized to $\ket{\psi(0)} = (\ket{E_1} + \ket{E_3})/\sqrt{2}$ in the presence of an in-plane field, the spin–state-dependent Larmor precession axis and frequency generate AC noise that manifests as the coherence modulation function, $Q_i(T)$, representing the back-action on the sensor. From a quantum-mechanical perspective, this corresponds to the entanglement dynamics between the electronic spin and the $i$-th nuclear spin.

With the analytical fit function established, the experimentally measured spin-echo profiles are fitted to Eq.~(\ref{eq:ESEEM}), enabling the extraction of the hyperfine components, $A_{\mu\nu}^{(i)}$, for all three nuclear spins ($i = 1, 2, 3$), together with the phenomenological decay constant, $T_{1/e}$, the stretch factor, $\beta$, and the contrast term, $c$. There are seven free parameters to be determined, $\{ A^{(1)}_{xx}, A^{(1)}_{yy}, \phi,\mathcal{E}_1, c, T_{1/e}, \beta\}$, constrained by the symmetries of the 2D hBN lattice and the fitted value of $\mathcal{E} = \sqrt{\mathcal{E}_1^2+\mathcal{E}_2^2} = 2\pi \times 18.5$~MHz. Among these, the hyperfine parameters $\{ A^{(1)}_{xx}, A^{(1)}_{yy} \}$, in-plane field orientation angle, $\phi$, and $\mathcal{E}_1$ are treated as global variables, shared across all measurements, while the phenomenological parameters $\{ c, T_{1/e}, \beta \}$ are treated as local variables, allowed to vary for each measurement to capture angle-dependent decoherence effects (Extended Data Fig.~\ref{fig:hyperfine}, Supplementary Information). We use a numerical optimization algorithm based on differential evolution to minimize the mean squared error between the measured and predicted spin-echo time traces across all external field angles, and the resulting best-fit hyperfine parameters are summarized in Fig.~\ref{fig:Fig2}h of the main text. A comparison showing fits with fixed hyperfine values from Refs. \cite{ivady2020ab, gong2024isotope} is shown in Extended Data Fig.~\ref{fig:compare}.

\subsection{Analysis of decoherence profiles}
In Fig.~\ref{fig:Fig3} of the main text, sensor decoherence profiles are measured under two different external field orientations, using the CPMG pulse sequence with increasing number of $\pi$ pulses, $N$, to achieve more effective decoupling of environmental noise sources. To quantify coherence extension under dynamical decoupling, each experimentally measured decoherence profile is fit to
\begin{align}
    C_N(T) = \tilde{C}_N(T) \exp \left[ -\left(\frac{T}{T_{2,N}}\right)^{\beta_N} \right],
\end{align}
where $\tilde{C}_N(T)$ captures the envelope modulation of the $N$-pulse CPMG sequence, and $T_{2,N}$ and $\beta_N$ denote the $1/e$ coherence time and stretch exponent, respectively, with the subscript $N$ explicitly indicating their dependence on the pulse number. The $T_{2,N}$ scaling result is shown in Fig.~\ref{fig:Fig3}d of the main text, with $\beta_N = 1.5 \pm 0.2$ exhibiting no noticeable dependence on $N$.

The envelope modulation function, $\tilde{C}_N(T)$, is modeled differently for the two field configurations: for the $B_z$ case with $\theta = 0^\circ$, no coherence modulation is expected due to the complete decoupling of the axial hyperfine interactions, so $\tilde{C}_N(T) = 1$ at all interrogation times $T$. In contrast, for the $B_\perp$ case with $\theta = 90^\circ$, $\tilde{C}_N(T) = \left[(1-c_N) + c_N \prod_{i=1}^3 Q_i(T/N)\right]$, where $c_N$ is the $N$-dependent phenomenological modulation contrast, and the spin-echo modulation function, $Q_i(T/N)$, is evaluated at $T/N$, reflecting that an $N$-pulse CPMG sequence consists of $N$ repetitions of the spin-echo sequence, each of duration $T/N$. We speculate that the phenomenological modulation constant, $c_N$, arises from inhomogeneity in hyperfine interaction strengths within the ensemble sensor or from additional contributions of slower coherent hyperfine interactions with more distant nuclear spins in the bath.

\subsection{Accurate noise PSD reconstruction}
The filter function formalism enables reconstruction of the noise PSD, $S(\omega)$, by solving the inversion problem in Eq.~(\ref{eq:chi(T)}), which relates the noise spectrum to the experimentally measured coherence, $C_N(T)$, through the filter function $F_N(\omega, T)$. To facilitate inversion, the filter function is often approximated as a Dirac delta function peaked at the principal resonance frequency $\omega = \pi N/T$. This approximation, however, becomes inaccurate when the finite duration of the $\pi$ pulses, $t_\pi$, is not negligible compared to the free-evolution interval, $\tau$, between adjacent pulses, or when $N$ is relatively small. To address this, we employ the {\it exact} form of the filter function~\cite{ishikawa2018influence}, which explicitly incorporates finite pulse-duration and finite-$N$ effects:
\begin{align}
\begin{split}
F_{N}(\omega,T) &=\Big|1+(-1)^{N+1}e^{i\omega T}\\
&+2\sum_{k=1}^N(-1)^ke^{i\omega t_{k}}\cos\left(\omega t_{\pi}/2\right)\Big|^{2},
\end{split}
\end{align}
where $t_k$ denotes the time corresponding to the center of the $k$-th pulse. Extended Data Fig.~\ref{fig:filter} shows the spectral profiles of the exact filter functions for the values of $N$ used in this work.

Crucially, when using the exact filter function, reconstruction of the noise PSD requires evaluating the frequency-domain integral over all frequencies (see the right-hand side of Eq.~(\ref{eq:chi(T)})). To this end, we approximate the continuous integral by a discretized Riemann sum using the trapezoidal method, with a maximum frequency cutoff of $\omega_\text{max}/2\pi = 100$~MHz and $\approx$$10^5$ logarithmically sampled frequency points, guided by the experimentally measured coherence times. Furthermore, to account for the presence of multiple sharp resonances whose widths scale as $1/N$, we introduce more densely spaced sampling points around these resonances to ensure convergence of the numerical integral.

In principle, to enable a model-free reconstruction of the noise PSD, $S(\omega)$ may be represented as a high dimensional vector sampled on the same frequency grid employed for the numerical integration. However, the resulting system of linear equations is underdetermined due to greater numerical sampling in frequency space than experimental time points. Instead, since the measured $C_N(T)$ profiles exhibit monotonically decaying behaviors, particularly in the magnetic-field sensing mode, we adopt a $1/\omega^\alpha$-noise ansatz for the noise PSD, $S(\omega) = S_0 / \omega^\alpha$, characterized by only two parameters, $\{S_0, \alpha\}$. Numerical optimization is performed using a differential evolution algorithm, followed by a final round of local gradient descent, to independently extract the best-fit parameters for each decoherence profile, $C_N(T)$, measured with different $N \in \{ 1, 8, 128, 256, 512 \}$. The results of this optimization yield a piecewise reconstruction of the noise PSD. For each $C_N(T)$ profile, the piecewise PSD is computed only within the statistically reliable region, determined by $\sigma_N < C_N(T) < 1 - \sigma_N$, where $\sigma_N$ is the standard deviation of the measured coherence signal in the time domain. The corresponding best-fit noise parameters are compiled in Table~\ref{tab:C(t)_fit_params}.

\renewcommand{\arraystretch}{1.2} 
\begin{table}[h!]
\centering
\begin{tabular}{c|c|c|c|c|c}
\hline
$N$ & 1 & 8 & 128 & 256 & 512 \\
\hline
\rule{0pt}{2.6ex} \textbf{$S_0$} 
    & $1.92 \times 10^{6}$ 
    & $7.04 \times 10^{5}$ 
    & $8 \times 10^{31}$ 
    & $2.83 \times 10^{9}$ 
    & $7.87 \times 10^{25}$ \\
$\alpha$ 
    & 0.73 
    & 0.69 
    & 5.20 
    & 1.24 
    & 3.39 \\
\hline
\end{tabular}
\caption{Extracted noise parameters from individual $C_N(T)$ fits based on a $1/\omega^\alpha$ noise model.}
\label{tab:C(t)_fit_params}
\end{table}

The piecewise noise PSD probes distinct frequency intervals set by the $N$-dependent resonances of the filter function. Since these individual PSDs originate from a common underlying spectrum, we construct a composite noise PSD using the same optimization procedure as above, guided by an empirical fit to the extracted profiles. Specifically, the fit function for the global composite noise PSD is defined as
\begin{align}
    S(\omega) = \frac{S_0}{\omega^\alpha\left[1+\left(\omega/\omega_c\right)^{\beta -\alpha}\right]}, 
\end{align}
where $\beta \geq \alpha$ is assumed, and $\omega_c$ denotes the crossover frequency. By construction, the spectrum reduces to $1/\omega^\alpha$ in the low-frequency limit ($\omega \ll \omega_c$) and to $1/\omega^\beta$ in the high-frequency limit ($\omega \gg \omega_c$), as illustrated by the solid line in Fig.~\ref{fig:Fig4}b of the main text. The corresponding best-fit parameters and their uncertainties are compiled in Table~\ref{tab:combined_noise_params}.

\renewcommand{\arraystretch}{1.2} 
\begin{table}[h!]
\centering
\begin{tabular}{l|c|c}
\hline
Parameter & Value & Uncertainty \\
\hline
\rule{0pt}{2.6ex}$S_0$             & $4.82 \times 10^{6}$ & $1.55 \times 10^{6}$ \\
$\alpha$                            & 0.79                 & 0.11 \\
$\beta$                             & 6.13                & 0.95\\
$\omega_c/2\pi$ (Hz)                & $6.24 \times 10^{6}$ & $3.84 \times 10^{5}$ \\
\hline
\end{tabular}
\caption{Extracted noise parameters for the global composite noise PSD derived from piecewise PSDs.}
\label{tab:combined_noise_params}
\end{table}

Note that all noise PSDs presented in Fig.~\ref{fig:Fig4} of the main text are plotted in units of linear frequency $f$, rather than angular frequency $\omega$, using the conversion $S(f) = 2\pi S(\omega)$ and $f = \omega / 2\pi$. The code base for implementing the filter function and optimization can be found at \texttt{https://github.com/nonohuff/FWDD}. 

\subsection{Numerical simulations of dynamical decoupling} \label{sec:sim_DD}
Extended Data Fig.~\ref{fig:decoupling} presents numerically simulated coherence times for $N$-pulse CPMG and XY8 sequences under magnetic-field noise, compared with the experimentally measured values. For the numerical simulations, we solve the time-dependent Schr\"{o}dinger equation using QuTiP, with noise statistics determined by the reconstructed magnetic-noise PSD (Fig.~\ref{fig:Fig4}b). Specifically, we consider the following time-dependent {\it stochastic} Hamiltonian for an effective two-level system spanned by $\{ \ket{\uparrow}, \ket{\downarrow} \}$,
\begin{align}
    \hat{H}(t) = \frac{1}{2}\left[ \Omega_x (t) \hat{\sigma}_x + \Omega_y (t) \hat{\sigma}_y \right] - \frac{1}{2} \Delta(t) \hat{\sigma}_z, \label{eq:H_DD}
\end{align}
where $\hat{\sigma}_{x,y,z}$ are the Pauli operators of the two-level system, $\Delta(t)$ is a random detuning-noise time trace generated via Monte Carlo sampling from the inverse Fourier transform of the reconstructed PSD, and $\Omega_{x,y}(t)$ are the time-dependent Rabi frequencies defined by the control pulse sequence, which accounts for finite pulse-duration effects. For example, $\Omega_x(t) = \Omega_0$ or $\Omega_y(t) = \Omega_0$ when control pulses are applied for a duration $t_\pi$ at a fixed bare Rabi frequency $\Omega_0$, with the $x/y$ components determined by the phase of the $\pi$ pulse; otherwise, $\Omega_{x,y}(t) = 0$ during the free-evolution intervals between adjacent pulses. The detuning noise, however, remains on at all times.

Starting from the initial superposition state $\ket{\psi(0)} = (\ket{\uparrow} + \ket{\downarrow})/\sqrt{2}$, we numerically compute the coherence, $C(t) = \langle \psi(t) | \hat{\sigma}_x | \psi(t) \rangle$, by sweeping the interrogation time from $t = N t_\pi$ to $t = T$, where $\ket{\psi(t)}$ is the state evolved under a noisy Hamiltonian $\hat{H}(t)$ with an $N$-pulse control sequence. This procedure is repeated for 100 Monte Carlo realizations to generate 100 instances of $C(t)$, from which we obtain the average decoherence profile, $\overline{C}(t)$, and extract the $1/e$ coherence time $T_2$. The entire process is also repeated for varying number of $\pi$ pulses, $N$, to probe the scaling of coherence time as a function of $N$.

As shown in Extended Data Fig.~\ref{fig:decoupling}, our numerical simulations are in good agreement with the experimental results for both CPMG and XY8 sequences. We find that the $\pi$-pulse phases play a crucial role: the CPMG sequence, with all $\pi$ pulses applied along the same axis, yields longer coherence times than XY8. We attribute this to spin-locking–like protection in the high microwave duty-cycle regime when finite pulse-duration effects are included, since a spin initially polarized along the $x$-axis remains stationary under the rapid application of many finite-duration $\pi$ pulses along the same axis~\cite{rizzato2023extending}.

\subsection{Numerical simulations of sensing performance}
In ensemble-based quantum sensing, performance can be limited by disorder (inhomogeneous broadening) in qubit transition frequencies as well as by control imperfections such as rotation angle errors and finite pulse durations. Using numerical simulations, we investigate their impact on sensing a particular AC signal under the CPMG and XY8 sequences, as shown in Extended Data Fig.~\ref{fig:robustness}.

\textit{Disorder effect:} Similar to the numerical procedure introduced in Sec.~\ref{sec:sim_DD}, we consider an effective two-level system with the Hamiltonian,
\begin{align}
    \hat{H}(t) = \frac{1}{2}\left[ \Omega_x (t) \hat{\sigma}_x + \Omega_y (t) \hat{\sigma}_y \right] - \frac{1}{2} \Delta \hat{\sigma}_z.
\end{align}
Here, the qubit detuning $\Delta$ is assumed to be purely dominated by static inhomogeneous broadening without additional time-dependent noise, and $\Omega_{x,y}(t)$ are defined by the control pulse sequence. We then include an additional Hamiltonian term, $\hat{H}_\text{ac}(t)$, which captures the interaction between the sensor spin and the AC signal:
\begin{align}
\hat{H}_\text{ac}(t) = \frac{1}{2} S_\text{ac}\sin(\omega_\text{ac} t + \phi_\text{ac}) \hat{\sigma}_z,
\end{align}
where $S_\text{ac}$, $\omega_\text{ac} = 2\pi/T_\text{ac}$, and $\phi_\text{ac}$ denote the amplitude, angular frequency, and phase of the target AC noise, respectively, with $\phi_\text{ac}$ set to 0 for simplicity. To detect the AC signal resonantly, we synchronize the control sequence, $\Omega_{x,y}(t)$, with the AC signal by imposing the resonance condition. The time-dependent Schr\"{o}dinger equation is then solved using the total Hamiltonian, $\hat{H}_\text{total}(t) = \hat{H}(t) + \hat{H}_\text{ac}(t)$, to numerically calculate the time-evolved state, $\ket{\psi(t)}$, starting from the initial superposition state $\ket{\psi(0)} = (\ket{\uparrow} + \ket{\downarrow})/\sqrt{2}$. 

In Extended Data Figs.~\ref{fig:robustness}b,c, we present the coherence $C(t) = \langle \psi(t) | \hat{\sigma}_x | \psi(t) \rangle$ as a function of time for different detuning values, $\Delta$, and compare the sensing performance of CPMG and XY8 sequences with $N=128$ pulses. These simulations show how detuned spin sensors across a wide range of disorder values respond to different sensing sequences, highlighting the robustness of these sequences in ensemble-based quantum sensing. Signal detection is indicated by a reduction in coherence, and the time axis is normalized such that a value of 1 corresponds to the principal resonance condition. Additional details are provided in the figure caption.

\textit{Control error effect:} To achieve both enhanced sensitivity and spectral selectivity, it is desirable to apply many $\pi$ pulses, since the spectral width of the filter function’s resonance scales as $1/N$. Consequently, minimizing control imperfections is essential, as over- or under-rotations in individual pulses can accumulate in large-$N$ sensing sequences, leading to additional decoherence and ultimately limiting sensitivity. To examine this effect, we consider the $N$-pulse control unitary operator, $\hat{U}_\epsilon(N)$, to analyze the robustness of each sensing sequence to control imperfections arising from rotation-angle errors, $\epsilon$, in the applied pulses. The unitary operator for the 8-pulse CPMG sequence is given by $\hat{U}_\epsilon(8) = \hat{U}_{\epsilon,x}^8$, while that for the XY8 sequence is given by $\hat{U}_\epsilon(8) = \hat{U}_{\epsilon,x} \hat{U}_{\epsilon,y} \hat{U}_{\epsilon,x} \hat{U}_{\epsilon,y} \hat{U}_{\epsilon,y} \hat{U}_{\epsilon,x} \hat{U}_{\epsilon,y} \hat{U}_{\epsilon,x}$, with $\hat{U}_{\epsilon,x} = e^{-i(\pi+\epsilon) \hat{\sigma}_x/2}$ and $\hat{U}_{\epsilon,y} = e^{-i(\pi+\epsilon) \hat{\sigma}_y/2}$. Accordingly, the $N$-pulse operator is defined as $\hat{U}_\epsilon(N) = [\hat{U}_\epsilon(8)]^{N/8}$ for both sequences, which yields the final state after $N$ pulses as $\ket{\psi(N)} = \hat{U}_\epsilon(N)\ket{\psi(0)}$, with the initial state $\ket{\psi(0)} = (\ket{\uparrow} + \ket{\downarrow})/\sqrt{2}$.

In Extended Data Figs.~\ref{fig:robustness}e,f, we present the $N$-pulse sequence fidelity, $F(N) = |\langle \psi(0) | \psi(N) \rangle|^2$, as a function of $N$ for different $\epsilon$ values and compare the robustness of CPMG and XY8 sequences. Ideally, $F(N)$ should remain close to 1 up to a tolerable error strength over many pulses, which is characteristic of self-correcting control sequences such as XY8. Additional details and discussion are provided in the figure caption.

\textit{Ensemble quantum sensing:} Having analyzed the effects of disorder and control errors separately, we identify XY8 as a robust pulse sequence for AC field sensing. Since our sample consists of a non-interacting ensemble of disordered spins, we further investigate how inhomogeneous broadening influences the $N$-pulse sequence fidelity. To this end, we define the disorder-averaged sequence fidelity, $\overline{F}(N) = \int_{-\infty}^\infty d\Delta \; \mathcal{N}_{\mu=0,\sigma=W}(\Delta) F(N, \Delta)$, where $\mathcal{N}_{\mu=0,\sigma=W}(\Delta)$ is the normal distribution of detuning $\Delta$ with mean $0$ and standard deviation $W$, and $F(N,\Delta)$ is the fidelity of preserving the initial coherence for a sensor detuned by $\Delta$ after $N$ pulses. As shown in Extended Data Fig.~\ref{fig:robustness}h, large-$N$ XY8 sequences remain robust against detuning disorder in an inhomogeneous sensor ensemble, provided the disorder strength $W$ is not too large compared with the control Rabi frequency, i.e., $W \lesssim 0.3\Omega$. Thanks to the built-in robustness of XY8, we numerically confirm that an ensemble sensor subjected to a 128-pulse XY8 sequence can successfully detect a target AC signal with performance nearly identical to the disorder-free ($W=0$) case, whereas CPMG exhibits much reduced sensing contrast (Extended Data Fig.~\ref{fig:robustness}i).

\subsection{Theoretical noise models}

The noise environment of boron-vacancy defects in hBN is expected to be highly heterogeneous, similar to that of other solid-state defects in close proximity to surfaces. From the reconstructed noise PSD, which exhibits a power-law decay, we infer three primary noise sources.

The first noise source is spin diffusion in the nuclear spin bath, leading to a frequency dependence of $S(\omega) \sim 1/\omega^{\alpha}$, where $\alpha < 1$~\cite{kubo1962resonance, de2009electron}. In our sample, the  $^{15}$N and $^{10}$B nuclear spins are dipolar-coupled, and their nuclear spin polarization can diffuse within the hBN crystal. As in Ref.~\cite{kubo1962resonance}, we treat the $V_B^-$ transition frequency as a stochastic variable that fluctuates due to its diffusive environment. The Fourier transform of the autocorrelation function of these fluctuations yields a power-law behavior with exponent $\alpha$, which depends on the system dimensionality~\cite{de2009electron}. Given the quasi-2D nature of our hBN sample, $\alpha$ is expected to range between 0.5 and 0.8~\cite{de2009electron}. The prediction from the diffusion-driven noise model is consistent with previous work reporting stretched-exponential coherence decay in $V_B^-$ ensemble measurements~\cite{haykal2022decoherence}.

The second noise source is an ensemble of fluctuating two-level systems (TLSs) coupled to the $V_B^-$ transition, giving rise to $1/\omega$ scaling~\cite{dutta1981low,mcwhorter1957semiconductor}. These fluctuators may originate from multiple sources, including charge traps at the hBN-substrate interface, surface molecular adsorbates, carbon impurities, or other spin defects in the lattice. Assuming that individual TLSs produce Lorentzian spectra with $1/\omega^2$ decay at high frequencies, and that defect activation energies follow a uniform distribution, a broad distribution of relaxation times can give rise to an overall $1/\omega$ spectrum~\cite{dutta1981low,mcwhorter1957semiconductor}.

The third noise source is a localized, Markovian cluster of interacting TLS defects, leading to a Lorentzian-like frequency dependence of $1/\omega^2$~\cite{machlup1954noise}. For example, this scenario can arise from localized charge traps near the $V_B^-$, structural defects associated with folds or imperfect edges in the 2D sheet, or nearby electronic spin impurities. These sources produce random telegraph noise, resulting in a Lorentzian PSD that decays as $1/\omega^2$ at frequencies above the switching rate of the fluctuating TLSs. Clusters of defects may also exhibit correlated switching, leading to non-Gaussian noise profiles~\cite{galperin2006non} and abrupt spectral jumps~\cite{stern2019spectrally}.

\subsection{Analysis of AC-field sensitivity}
For a non-interacting ensemble of $N_e$ quantum sensors detecting a spatially correlated signal, the AC-field sensitivity (in units of T/$\sqrt{\text{Hz}}$) is given by~\cite{barry2020sensitivity}
\begin{align}
\eta_{\text{ac}} &= \frac{\pi}{2\gamma} \frac{e^{(T/T_{2})^\beta}}{\mathcal{C}\sqrt{N_{e}}} \frac{\sqrt{T_\text{init}+T+T_\text{read}}}{T},
\end{align}
where $\mathcal{C}$ is the single-sensor readout efficiency, $\gamma$ is the gyromagnetic ratio, and $T_\text{init}$ and $T_\text{read}$ denote the sensor initialization and readout times, respectively. In particular, $\mathcal{C}\sqrt{N_e}$ can be regarded as an ensemble readout efficiency, given by
\begin{align}
\frac{1}{\mathcal{C}\sqrt{N_{e}}} &= \frac{1}{\sqrt{N_e}} \sqrt{1 + \frac{2(\alpha_0+\alpha_{1})}{(\alpha_{0}-\alpha_{1})^2}} \approx \sqrt{\frac{2(\alpha_0'+\alpha_{1}')}{(\alpha_{0}'-\alpha_{1}')^2}},
\end{align} 
where $\alpha_0' = N_e\alpha_0$ and $\alpha_1' = N_e \alpha_1$ denote the number of photons collected during a single readout pulse of duration $T_\text{read}$ for the optically bright and dark states, respectively, with $\alpha_{0,1}$ being the mean photon counts from single spins~\cite{degen2017quantum}. In our experiment, the achievable maximum optical contrast is measured to be $2(\alpha_0' - \alpha_1')/(\alpha_0' + \alpha_1') \approx 0.1$, which yields an ensemble readout efficiency of $\mathcal{C}\sqrt{N_{e}} \approx 0.035$. Using $T_2 = 29~\mu\text{s}$ obtained from a 512-pulse XY8 sequence in the magnetic sensing mode, together with $T_\text{init} = 0.4~\mu\text{s}$, $T_\text{read} = 0.6~\mu\text{s}$, and $\gamma = \gamma_z$, we estimate the sensitivity of our boron-vacancy ensemble sensor to be $\eta_\text{ac} \approx 138$~nT/$\sqrt{\text{Hz}}$ at an optimal sensing time of $T+T_\text{init}+T_\text{read} \approx 10~\mu\text{s}$ (Extended Data Fig.~\ref{fig:sensitivity}). It is important to note that the AC sensitivity is estimated using XY8, as this sequence better preserves spectral information and offers higher sensitivity compared with CPMG (see Fig.~\ref{fig:Fig4} of the main text).

We note that the bright photon count per readout, $\alpha_0 \approx 0.36$, measured with a single-photon counting module, is limited by saturation and therefore does not fully harness the capability of ensemble sensing. This suggests significant room for sensitivity enhancement by employing a detector with a larger dynamic range (Extended Data Fig.~\ref{fig:sensitivity}), which we leave for future work.

\sisetup{
    separate-uncertainty = true,
    multi-part-units = single
}



\bibliography{main}

\newcounter{extfigure}
\renewcommand{\theextfigure}{\arabic{extfigure}}

\newcommand{\extcaption}[1]{%
  \refstepcounter{extfigure}%
  \par\noindent\textbf{Extended Data Fig.~\theextfigure.} \justifying #1\par
}

\begin{figure*}[htbp] 
\begin{center}
\includegraphics[scale=1]{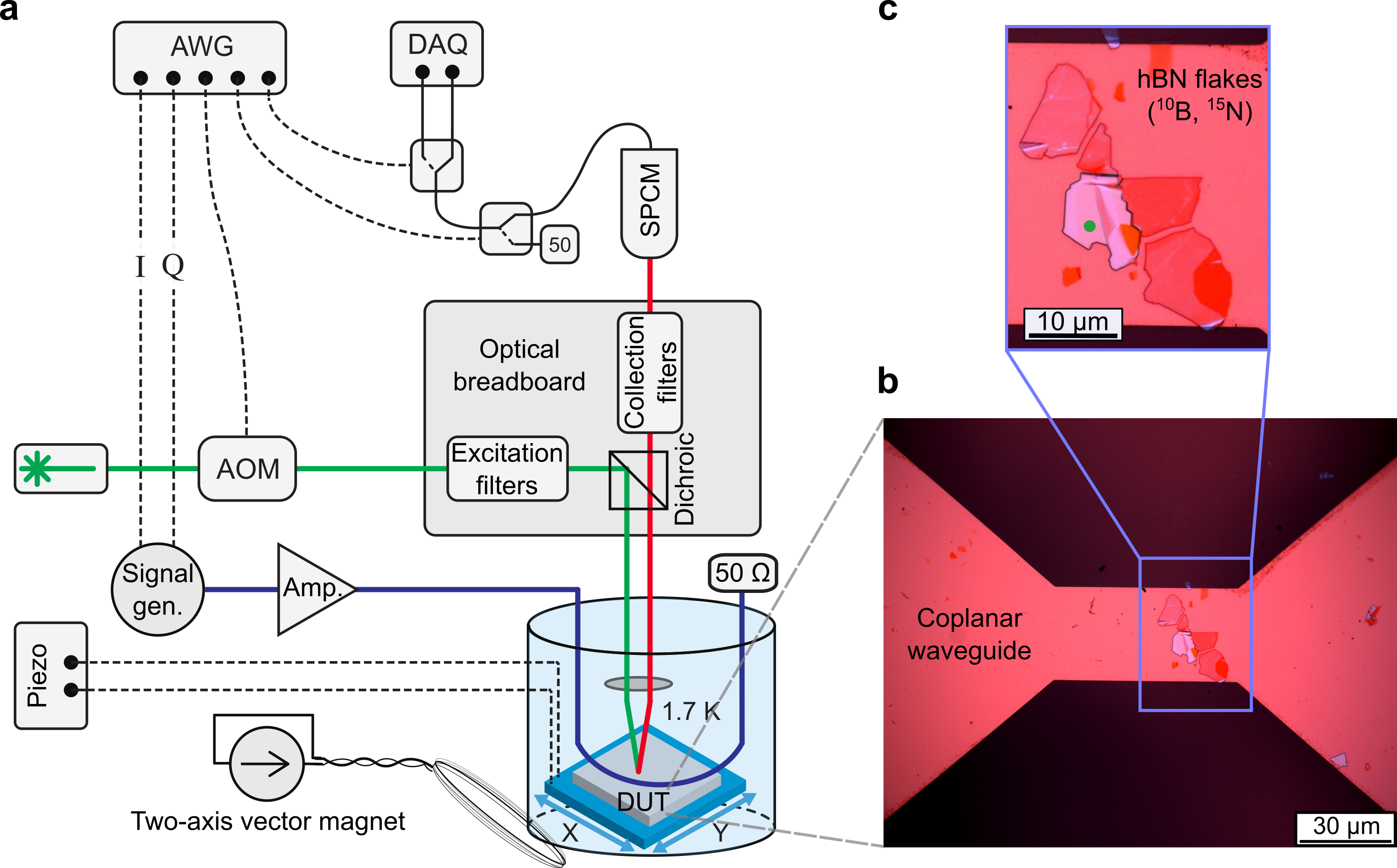}
\extcaption{\textbf{Experimental setup and device.}
\textbf{a,} Schematic of the experimental setup used to probe the hBN device under test (DUT). An arbitrary waveform generator (AWG) provides control signals to a signal generator for synthesizing microwave pulse sequences, drives an acousto-optic modulator (AOM), and defines and synchronizes signal acquisition with a single-photon counting module (SPCM). A 532-nm green laser is delivered to the cryostat and modulated by the AOM, while the resulting photon emission is separated by a dichroic filter and passed through a set of collection filters before detection (Methods). The sample position is precisely controlled with a piezo-actuated stage, and a two-axis vector magnet is employed to tune the eigenstates of the spin sensor. \textbf{b,} Optical microscope image of isotopically purified h$^{10}$B$^{15}$N flakes transferred onto a coplanar waveguide. \textbf{c,} Magnified view of the hBN flakes, with the green dot marking the measurement location corresponding to the data presented in this work.
}
\label{fig:setup}
\end{center}
\end{figure*}

\begin{figure*}[htbp] 
\begin{center}
\includegraphics[scale=1]{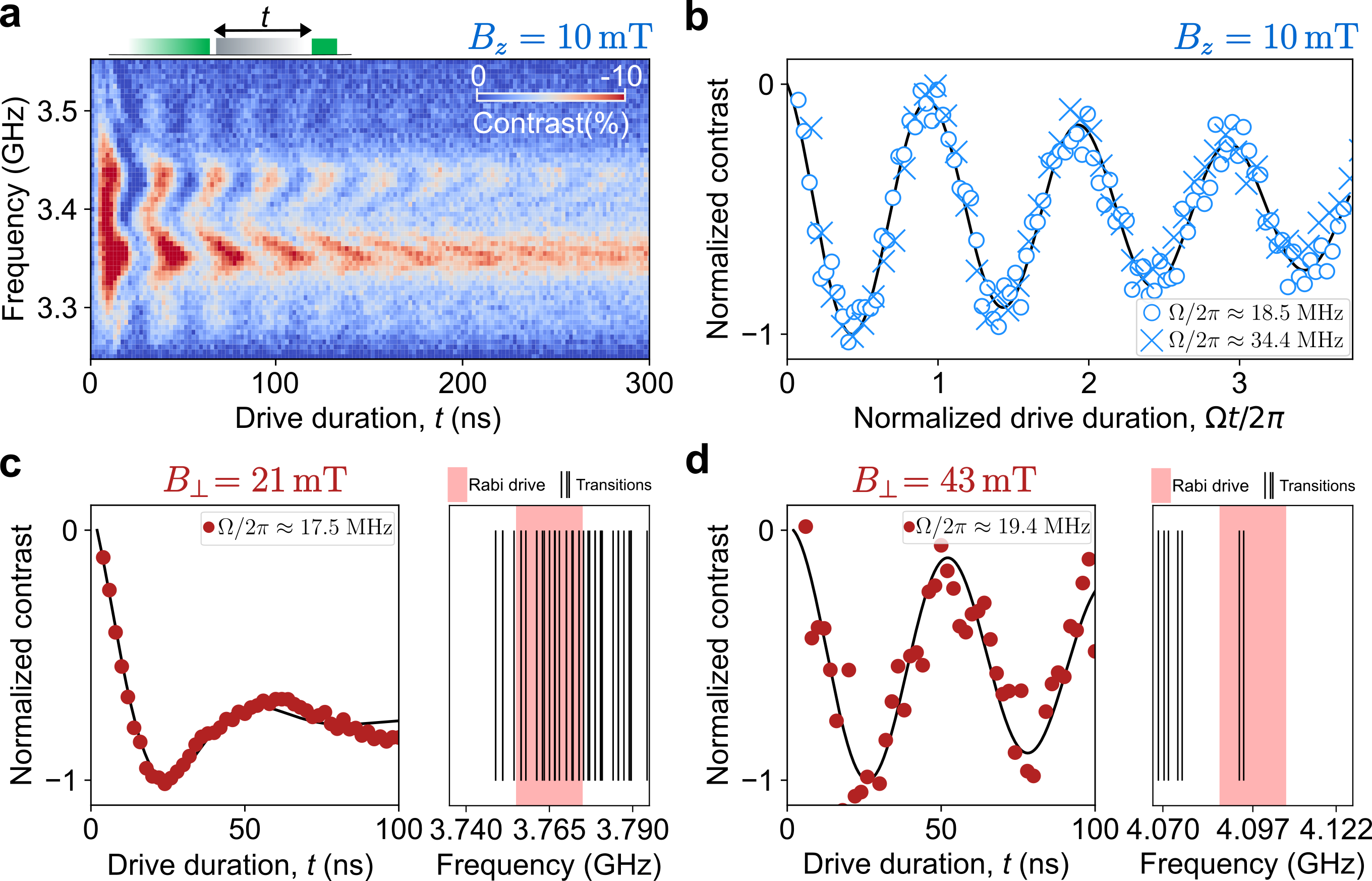}
\extcaption{\textbf{Rabi oscillations in a boron-vacancy spin ensemble.}
\textbf{a,} Coherent Rabi oscillation dynamics of a boron-vacancy spin ensemble in its lower spin resonance manifold, defined by the $\ket{0} \leftrightarrow \ket{-1}$ transition, under an out-of-plane field of $B_z = 10$ mT. After initializing the boron-vacancy spin ensemble into the $\ket{0}$ state via off-resonant green excitation, we apply a microwave drive at a fixed frequency but with variable duration $t$, inducing population exchange between the $\ket{0}$ and $\ket{-1}$ states (see the Rabi driving sequence at the top of the figure). Four distinct Rabi oscillations are observed at different microwave frequencies, corresponding to four hyperfine-split resonances arising from the axial hyperfine interaction with the three nearest-neighbor $^{15}$N nuclear spins (see main text for details). \textbf{b,} Horizontal line cut from \textbf{a} at a microwave frequency of $\approx 3.35$ GHz, showing Rabi oscillations with maximum contrast. Two sets of oscillations, driven at Rabi frequencies of $\Omega/2\pi \approx 18.5$ MHz and $\Omega/2\pi \approx 34.4$ MHz, are compared on the normalized time axis $\Omega t/2\pi$. The decay envelopes are found to be scale-invariant, indicating that Rabi decoherence arises from shot-to-shot fluctuations in the Rabi frequency. In all experiments, we used the lower Rabi frequency to minimize drive-induced heating. \textbf{c,} Rabi oscillations under an in-plane magnetic field of $B_{\perp} = 21~\mathrm{mT}$ at a Rabi frequency of $\Omega/2\pi \approx 17.5$ MHz (left). The barcode representation of spin resonance frequencies (right) reveals a cluster of hyperfine-split transitions near a microwave carrier frequency of $\approx 3.765$ GHz, arising from the mixing of axial and transverse hyperfine interaction components. The red box marks the effective spectral window within which transitions are strongly driven. Driving densely packed but spectrally detuned transitions is attributed to the rapid decay of the observed Rabi oscillations. \textbf{d,} Same as \textbf{c}, but under a higher in-plane magnetic field of $B_{\perp} = 43~\mathrm{mT}$ at a Rabi frequency of $\Omega/2\pi \approx 19.4$ MHz (left). A longer Rabi decay time is observed at this higher field, which we attribute to reduced spectral crowding near a microwave carrier frequency of $\approx 4.097$ GHz compared to the lower-field case; in this regime, only a couple of hyperfine transitions are resonantly driven.}
\label{fig:rabi}
\end{center}
\end{figure*}

\begin{figure*}[htbp] 
\includegraphics[scale=1]{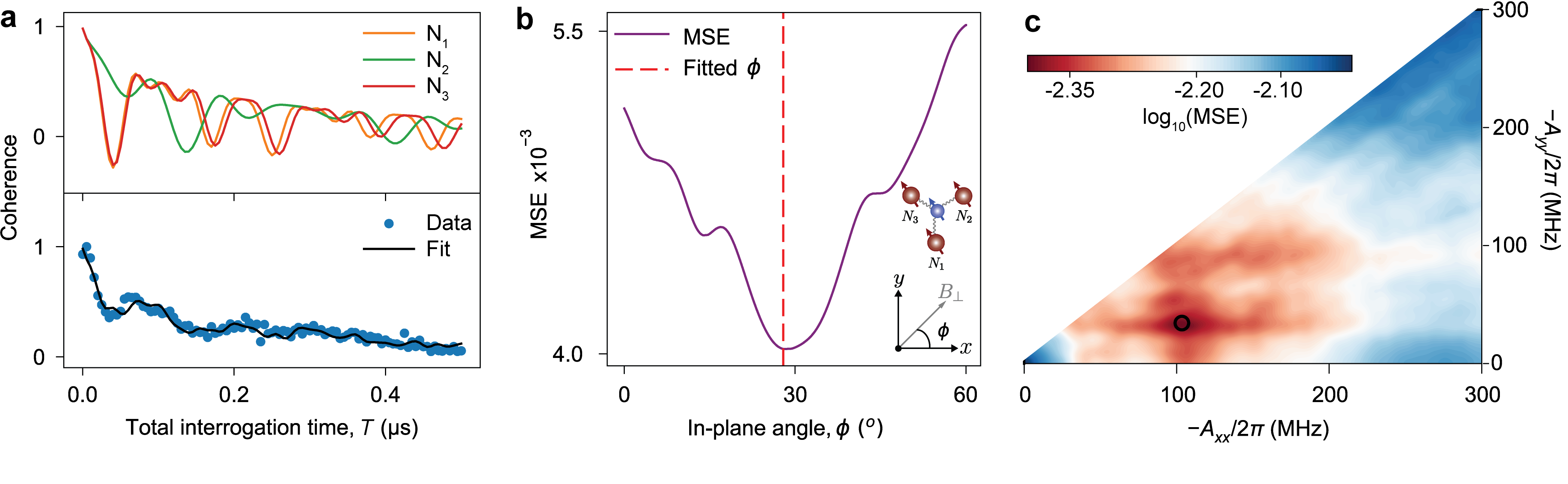}
\extcaption{
\textbf{Learning hyperfine Hamiltonian parameters via short-time spin-echo dynamics.} \textbf{a,} Coherence modulation of the central boron-vacancy spin under a spin-echo sequence, induced by hyperfine interactions with three nearest-neighbor $^{15}$N nuclear spins in the presence of an in-plane magnetic field $B_\perp$ ($\theta = 90^\circ$). The top plot shows the numerically simulated, site-resolved contributions from the individual nuclear spins, and the bottom plot shows the experimental results (same data as in Fig.~\ref{fig:Fig2}e) along with the theoretical prediction. The same phenomenological parameters $\{T_\text{1/e}, \beta\}$ obtained from the numerical optimization are applied to the simulation data to facilitate comparison (Supplementary Information). \textbf{b,} Extraction of the azimuthal angle, $\phi$, of an in-plane field, $B_\perp$, relative to the $x$-axis of the hBN lattice (inset). The orientation of the applied in-plane field yields varying axial and transverse projections along each nitrogen axis, resulting in a $\phi$-dependent coherence modulation profile. This angle sensitivity can be used to identify the most likely in-plane angle $\phi$ that best matches the experimental data. The mean square error (MSE) between the $\phi$-dependent model prediction and the experiment, plotted as a function of $\phi$, shows a minimum at the fitted value of $\phi \approx 28^\circ$ (vertical dashed line). \textbf{c,} Extraction of the transverse hyperfine interaction components from the first nitrogen atom (N$_1$), $\{A^{(1)}_{xx}, A^{(1)}_{yy}\}$. The MSE between the model prediction and the experiment is used to identify the optimal transverse hyperfine parameters, indicated by the black circular point on the heatmap. Note that the hyperfine parameter signs are chosen to be negative, following the DFT convention~\cite{ivady2020ab}.}
\label{fig:hyperfine}
\end{figure*}

\begin{figure*}[htbp] 
\includegraphics[scale=1]{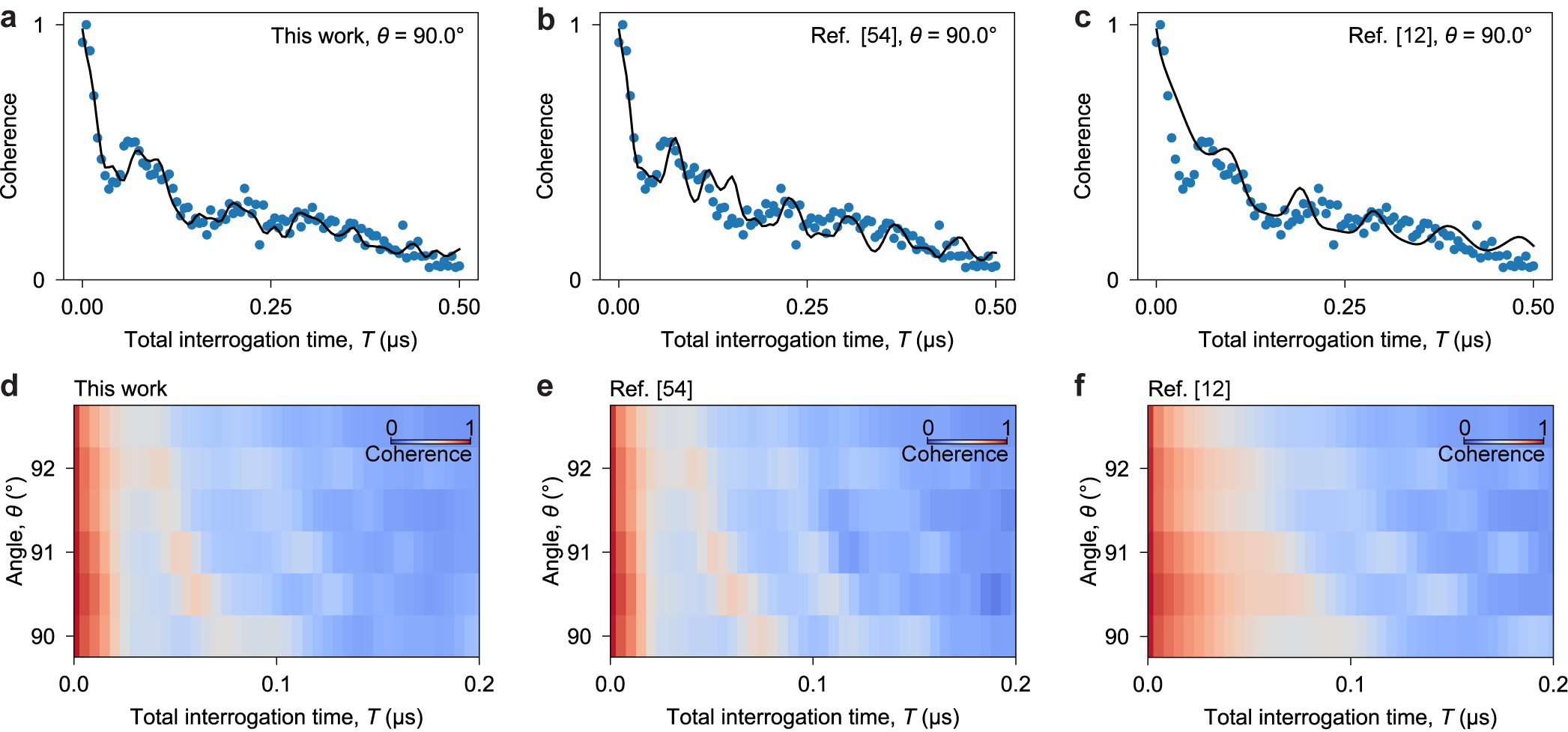}
\extcaption{
\textbf{Comparison of short-time spin-echo dynamics across different predictions.} \textbf{a–c,} Coherence modulations of the central boron-vacancy spin under a spin-echo sequence with an in-plane magnetic field of $B_\perp = 20$ mT, compared across three different reported sets of hyperfine interaction Hamiltonian parameters: our extracted values (see Fig.~\ref{fig:Fig2}h in the main text), Ref.~\cite{ivady2020ab}, and Ref.~\cite{gong2024isotope}. The black solid lines represent fits based on these models, while the blue markers denote the experimental data measured in this work. \textbf{d–f,} Sensitive variations in the simulated spin-echo dynamics as a function of the external field polar angle, $\theta$, relative to the $z$-axis. Each subplot is calculated using its respective reported set of hyperfine interaction parameters (Methods).
}
\label{fig:compare}
\end{figure*}

\begin{figure*}[htbp] 
\includegraphics[scale=1]{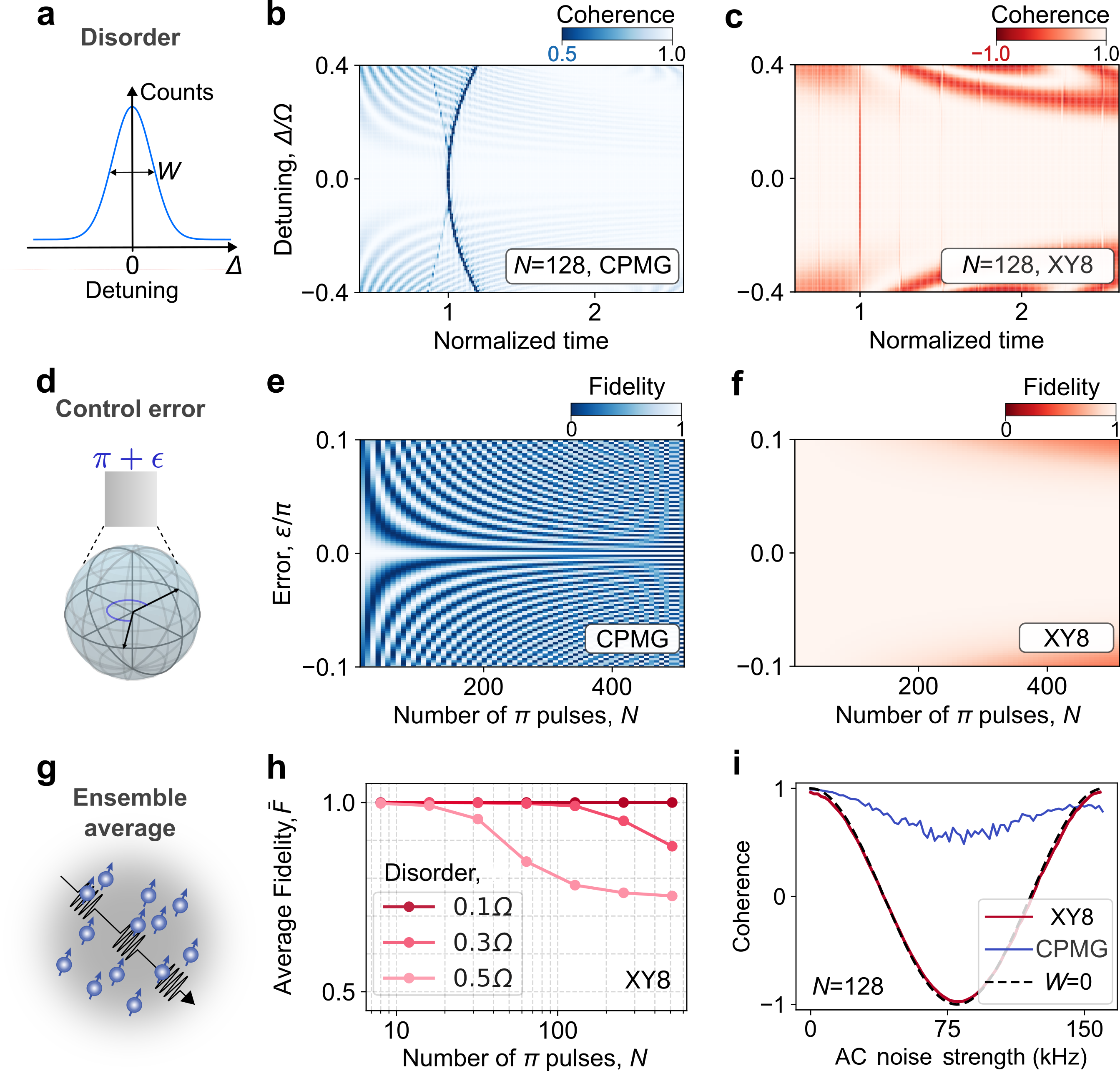}
\extcaption{\textbf{Robustness of different AC sensing sequences.} \textbf{a,} Schematic illustration of disorder in an ensemble of spin qubits, also referred to as inhomogeneous broadening, showing a normal distribution of qubit detuning values $\Delta$, with a full width at half maximum, $W$. \textbf{b, c,} Robustness to inhomogeneous broadening in detecting a target AC signal at the principal resonance condition (corresponding to the normalized time of 1) is shown for CPMG (\textbf{b}) and XY8 (\textbf{c}) sequences with $N=128$ pulses. For CPMG, where the periodic $\pi$ pulses share the same driving axis along the $x$-axis of the Bloch sphere, effective AC sensing is achieved only when the detuning, $\Delta$, is much smaller than the Rabi frequency, $\Omega$, of the $\pi$ pulse, i.e., when $|\Delta/\Omega| \ll 1$. For highly detuned sensors, resonances appear at shifted time points, giving rise to a systematic parabolic ``bowing'' of the resonance curve. In contrast, the $\pi$-pulse phases of XY8 are judiciously chosen to self-correct control imperfections, making the sequence robust against inhomogeneous broadening and preserving the principal resonance frequency across a wide range of detuning values. \textbf{d,} Schematic illustration of control imperfections in $\pi$ pulses, showing a rotation angle error of magnitude $\epsilon$. \textbf{e, f,} Robustness to rotation angle errors in preserving the initial quantum coherence is shown for CPMG (\textbf{e}) and XY8 (\textbf{f}) sequences with varying number of $\pi$ pulses, $N$. To quantify robustness, we define the $N$-pulse sequence fidelity as $F(N) = |\langle \psi(0)|\psi(N)\rangle|^2$. Here, $\ket{\psi(0)} = (\ket{\uparrow} + \ket{\downarrow})/\sqrt{2}$ is the initial equal superposition state with perfect coherence between $\ket{\uparrow}$ and $\ket{\downarrow}$, and $\ket{\psi(N)} = \hat{U}_\epsilon(N)\ket{\psi(0)}$ is the final state after applying $N$ \textit{imperfect} pulses with a rotation angle error $\epsilon$, characterized by the $N$-pulse unitary operator $\hat{U}_\epsilon(N)$ (Methods). Numerical simulations reveal that CPMG exhibits strong sensitivity to rotation-angle errors, whereas XY8 remains robust against control imperfections, demonstrating near-perfect fidelity for errors up to $|\epsilon/\pi| \sim 0.1$. \textbf{g,} Schematic illustration of global quantum control for ensemble sensors, where the sensing response is determined by the average behavior. \textbf{h,} Disorder-averaged fidelity as a function of $N$ for robust XY8 sequences with three different disorder strengths: $W = 0.1\Omega, 0.3\Omega,$ and $0.5\Omega$. See Methods for the definition of disorder-averaged fidelity. \textbf{i,} Robust ensemble-based AC sensing with an $N=128$ XY8 sequence, showing sinusoidal oscillations as a function of AC noise strength while exhibiting a sensing contrast comparable to the ideal disorder-free ($W=0$) case. In contrast, the CPMG sequence exhibits much lower contrast, highlighting the importance of robust control sequences in ensemble-based sensing.}
\label{fig:robustness}
\end{figure*}

\begin{figure*}[htbp] 
\includegraphics[scale=1]{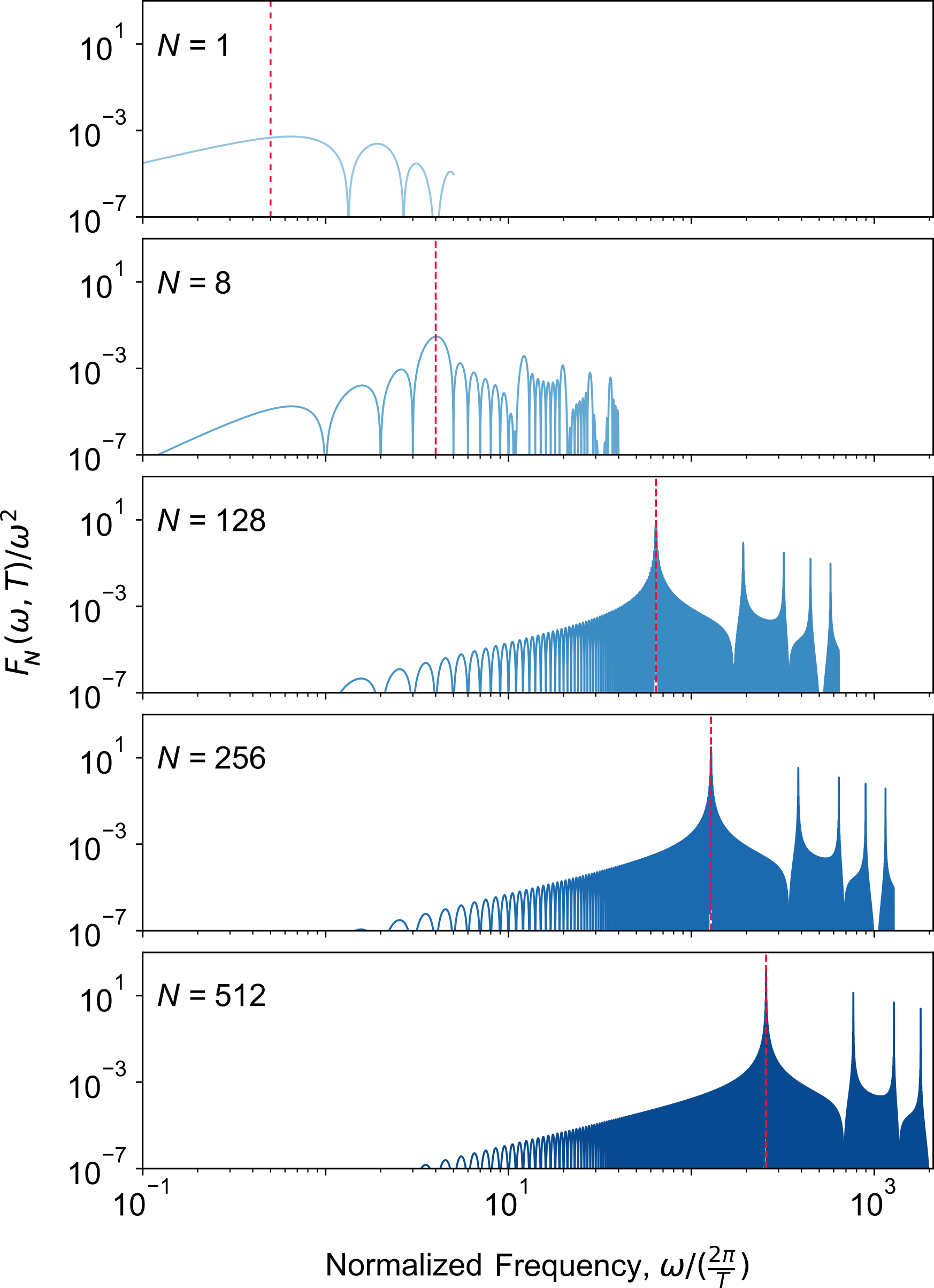}
\extcaption{
\textbf{Filter functions with finite control pulse duration.}  
Filter functions, $F_N(\omega, T)/\omega^2$, for periodic control sequences consisting of equidistant $N$ $\pi$ pulses are shown for $N = 1, 8, 128, 256$, and $512$. In each subplot, the vertical dotted line marks the principal resonance at angular frequency $\omega = 2\pi f_\text{res}$, where $f_\text{res} = N/2T$ and $T=N(\tau+t_{\pi})$. The inter-pulse delay, $\tau$, and finite pulse duration, $t_\pi$, are both set to $\tau = t_\pi = 24~\text{ns}$. As $N$ increases, the principal resonance shifts to higher frequencies, thereby enhancing noise decoupling and extending coherence times. 
}
\label{fig:filter}
\end{figure*}

\begin{figure*}[htbp] 
\includegraphics[scale=0.45]{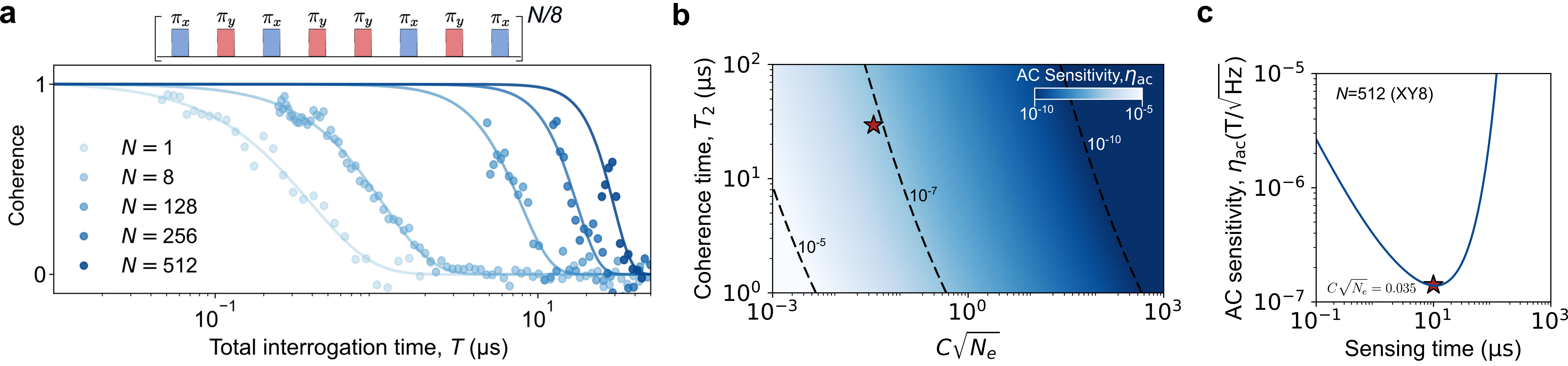}
\extcaption{ 
\textbf{AC magnetic field sensitivity analysis.} 
\textbf{a,} Coherence time traces measured with XY8 sensing sequences using varying number of $\pi$ pulses, $N$. The XY8 sequence consists of $\pi$ pulses with alternating phases, arranged in repeating trains of eight with mirror symmetry, as shown at the top. In the experiment, individual coherence profiles, $C_N(T)$, are fitted with a stretched exponential, $C_N(T) = e^{-(T/T_2)^{\beta}}$, where $T_2$ and $\beta$ denote the $N$-dependent coherence time and stretch exponent, respectively. The extracted $T_2$ values as a function of $N$ are reported in Fig.~\ref{fig:Fig4}c of the main text. \textbf{b,} AC magnetic field sensitivity, $\eta_\text{ac}$, as a function of coherence time, $T_2$, and effective readout efficiency of an {\it ensemble} sensor, $\mathcal{C}\sqrt{N_e}$. Here, $\mathcal{C}$ denotes the readout efficiency of a {\it single} sensor, and $N_e$ is the number of sensors used in ensemble-based sensing. The colormap is expressed in units of T/$\sqrt{\textrm{Hz}}$. The red star marks the estimated sensitivity based on experimentally characterized parameters, including an ensemble readout efficiency of $\mathcal{C}\sqrt{N_e} = 0.035$ (Methods). We expect that a more efficient detector, capable of fully harnessing the ensemble signal, would substantially improve sensitivity, potentially enabling nanotesla-level performance. \textbf{c,} Scaling of AC magnetic field sensitivity, $\eta_{\text{ac}}$, as a function of total sensing time for a 512-pulse XY8 sequence. Note that this sequence yields a coherence time of $T_2 \approx 29~\mu\mathrm{s}$ at a cryogenic temperature of $\approx$2K. For sensing times shorter than $T_2$, the sensitivity improves with increasing duration, while for times longer than $T_2$, decoherence limits performance and degrades sensitivity. This trade-off defines an optimal sensing time of $\approx 10~\mu\text{s}$, at which we estimate the sensitivity (red star). 
}
\label{fig:sensitivity}
\end{figure*}

\begin{figure*}[htbp] 
\includegraphics[scale=1]{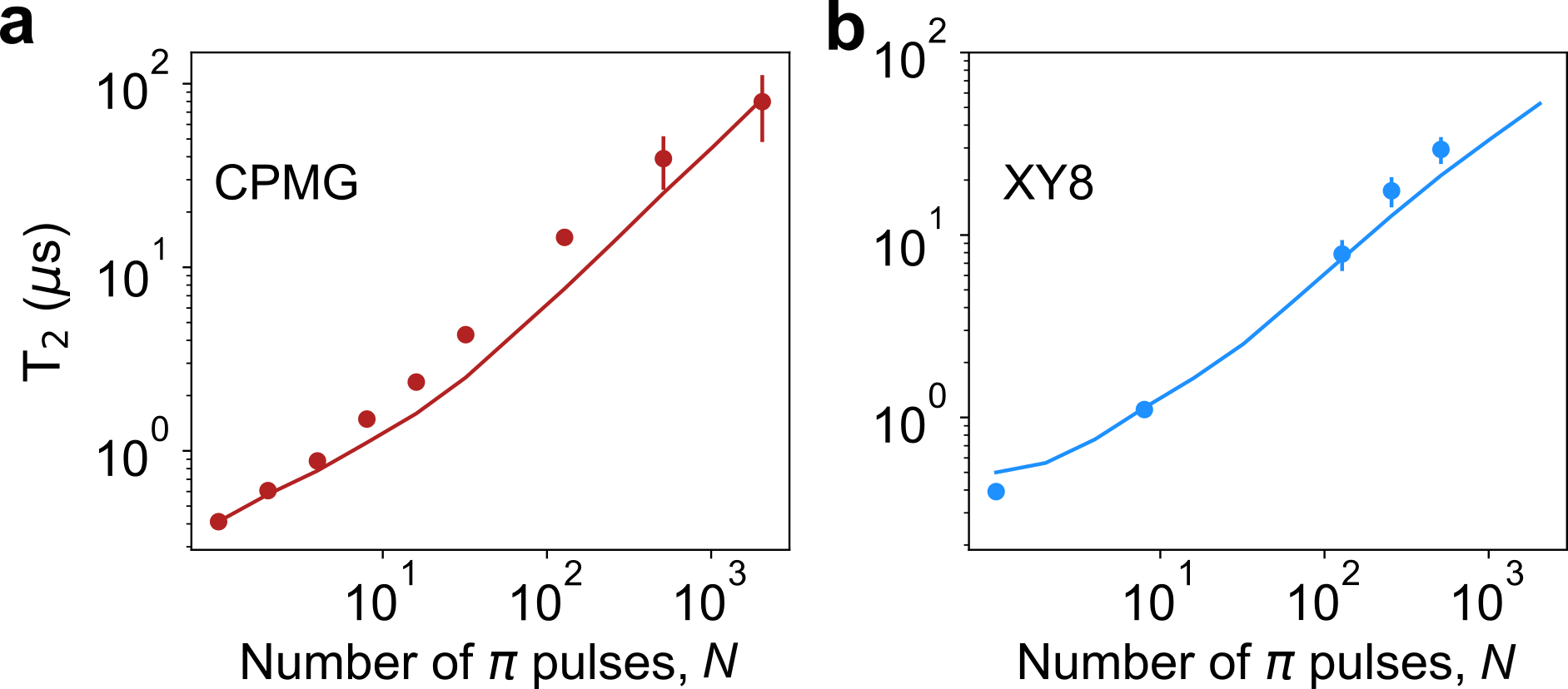}
\extcaption{
\textbf{Magnetic-noise decoupling performance of CPMG and XY8 sequences.} Experimentally measured coherence times, $T_2$, (markers) as a function of the number of $\pi$ pulses, $N$, are shown for \textbf{a,} CPMG and \textbf{b,} XY8 sequences. The solid lines represent numerical predictions obtained from full quantum-mechanical simulations, where the time-dependent Schr\"{o}dinger equation is solved under Monte Carlo–sampled frequency noise drawn from the learned noise PSD presented in Fig.~\ref{fig:Fig4}b of the main text (see Methods for simulation details). We find that the numerical simulation results are consistent with the experimental data for both sequences, corroborating the accuracy of our reconstructed magnetic-noise PSD. Notably, the coherence times under CPMG are longer than those under XY8, owing to the spin-locking effect, even though CPMG is not optimal for AC noise spectroscopy (Methods).
}
\label{fig:decoupling}
\end{figure*}

\clearpage
\onecolumngrid

\begin{center}
  {\Large \textbf{Supplementary Information for \\
  ``Quantum sensing with a spin ensemble in a van der Waals material''}\par}
  \vspace{0.5cm}
  {\normalsize
  Souvik Biswas,$^{1, *}$ 
  Giovanni Scuri,$^{1, *}$ 
  Noah Huffman,$^{2}$ 
  Eric I. Rosenthal,$^{1, 3}$ 
  Ruotian Gong,$^{4}$ 
  Thomas Poirier,$^{5}$ 
  Xingyu Gao,$^{6}$ 
  Sumukh Vaidya,$^{6}$ 
  Abigail J. Stein,$^{7}$ 
  Tsachy Weissman,$^{1}$ 
  James H. Edgar,$^{5}$ 
  Tongcang Li,$^{6,8}$ 
  Chong Zu,$^{4}$ 
  Jelena Vu\v{c}kovi\'{c},$^{1,}$\textsuperscript{†} 
  Joonhee Choi$^{1,}$\textsuperscript{‡} \par}
  \vspace{0.5cm}

  {\small
  $^{1}$\textit{Department of Electrical Engineering, Stanford University, Stanford, CA, USA} \\
  $^{2}$\textit{Department of Physics, Stanford University, Stanford, CA, USA} \\
  $^{3}$\textit{Present address: Sygaldry Technologies, Ann Arbor MI, USA}\\
  $^{4}$\textit{Department of Physics, Washington University, St. Louis, MO, USA} \\
  $^{5}$\textit{Tim Taylor Department of Chemical Engineering, Kansas State University, Manhattan, KS, USA} \\
  $^{6}$\textit{Department of Physics and Astronomy, Purdue University, West Lafayette, IN, USA} \\
  $^{7}$\textit{Department of Applied Physics, Stanford University, Stanford, CA, USA} \\
  $^{8}$\textit{Elmore Family School of Electrical and Computer Engineering, Purdue University, West Lafayette, IN, USA} \\
  \vspace{0.5cm}
  *These authors contributed equally to this work. \\
  \textsuperscript{†}\href{mailto:jela@stanford.edu}{jela@stanford.edu}, \\
  \textsuperscript{‡}\href{mailto:joonhee.choi@stanford.edu}{joonhee.choi@stanford.edu}
  }
\end{center}
\vspace{2cm}

\newcounter{supfigure}
\renewcommand{\thesupfigure}{\arabic{supfigure}}
\setcounter{table}{0}
\renewcommand{\thetable}{S\arabic{table}}
\setcounter{page}{1}
\renewcommand{\thepage}{SI \arabic{page}}

\newcommand{\supcaption}[1]{%
  \refstepcounter{supfigure}%
  \par\noindent\textbf{S.I. Fig.~\thesupfigure.} \justifying #1\par
}

\begin{figure*}[htbp] 
\includegraphics[scale=1]{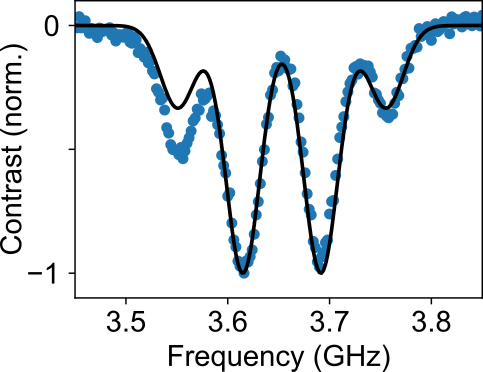}
\supcaption{
\textbf{Zero-field ODMR spectrum.} ODMR spectrum data (markers) measured under zero external magnetic field. In the absence of hyperfine interactions, the $\ket{0} \leftrightarrow \ket{-1}$ and $\ket{0} \leftrightarrow \ket{+1}$ transitions of the bare spin eigenstates are expected to be degenerate at $\approx$3.65~GHz, set by the zero-field splitting. However, the axial hyperfine interaction, $A_{zz}$, together with the strain term, $\mathcal{E}$, lifts this degeneracy and gives rise to a four-peak structure. Fitting the data with our model (black solid line) yields a strain strength of $\mathcal{E}/2\pi = 18.5 \pm 0.4~\mathrm{MHz}$.
}
\end{figure*}

\begin{figure*}[htbp] 
\includegraphics[scale=1]{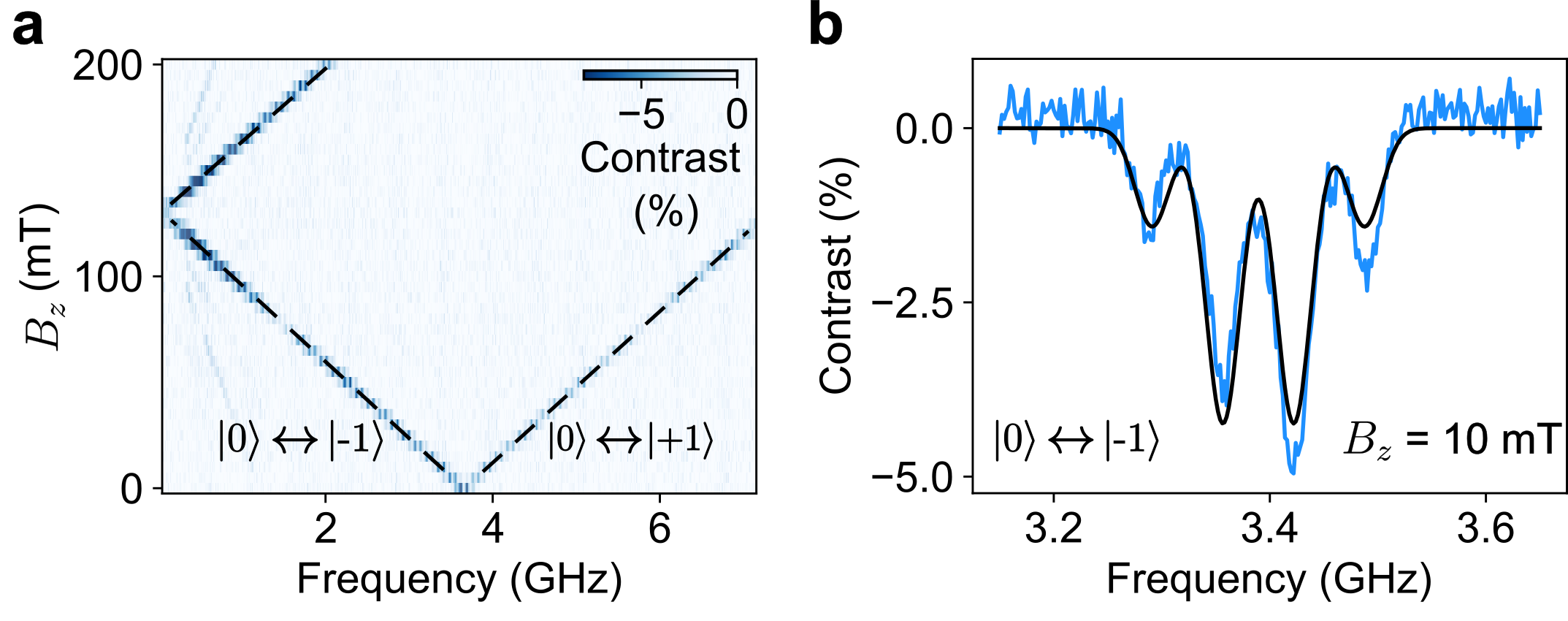}
\supcaption{
\textbf{ODMR spectroscopy under an axial magnetic field applied along the $z$-axis.} \textbf{a,} ODMR spectra as a function of the axial, out-of-plane magnetic field, $B_{z}$, showing two resonance branches that diverge linearly with field, corresponding to the $\ket{0} \leftrightarrow \ket{-1}$ and $\ket{0} \leftrightarrow \ket{+1}$ transitions. Black dashed lines represent fits to the resonance frequencies of each branch, yielding an out-of-plane gyromagnetic ratio of $\gamma_{z}/2\pi = 28 \pm 0.2~\mathrm{GHz/T}$, consistent with reported values~\cite{gong2024isotope}. \textbf{b,} Horizontal line cut of the ODMR spectrum at $B_{z} = 10~\textrm{mT}$ for the $\ket{0} \leftrightarrow \ket{-1}$ branch, showing four hyperfine-split resonances. The spectrum is fitted with a sum of four Lorentzian profiles with relative amplitudes of 1:3:3:1, from which we extract an axial hyperfine coupling strength of $A_{zz}/2\pi = -67 \pm 0.5~\textrm{MHz}$, with the negative sign following the DFT convention~\cite{ivady2020ab}.
}
\end{figure*}

\begin{figure*}[htbp] 
\includegraphics[width=0.82\textwidth]{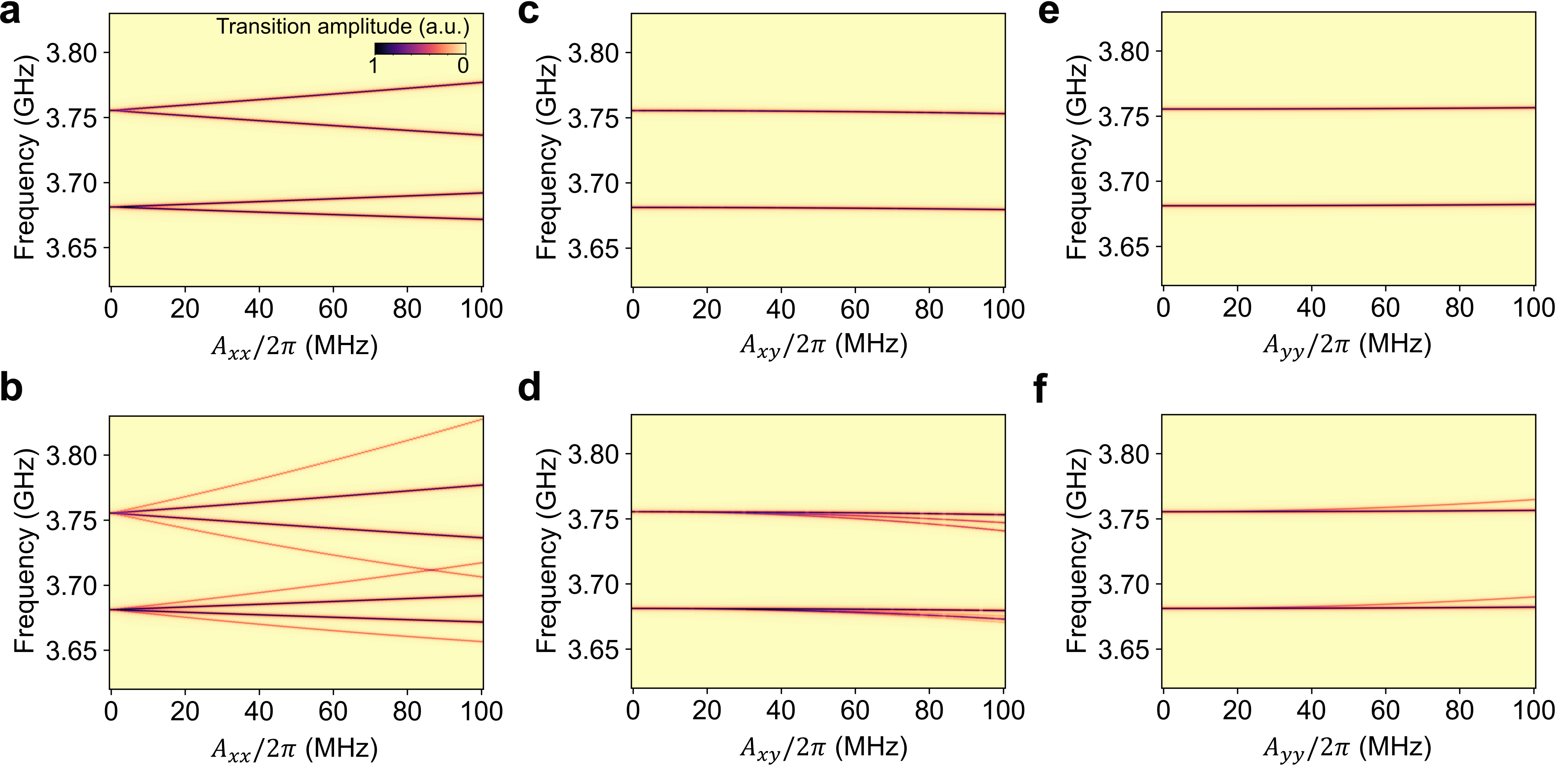}
\supcaption{
\textbf{Dependence of ODMR spectra on individual hyperfine parameters.} Perturbations of ODMR transition frequencies as a function of (\textbf{a, b}) $A_{xx}$, (\textbf{c, d}) $A_{xy}$, and (\textbf{e, f}) $A_{yy}$. In \textbf{a, c, e}, only a single nuclear spin is assumed to couple to the central spin for simplicity, whereas in \textbf{b, d, f}, three nuclear spins are coupled.
}
\end{figure*}

\begin{figure*}[htbp] 
\includegraphics[scale=1]{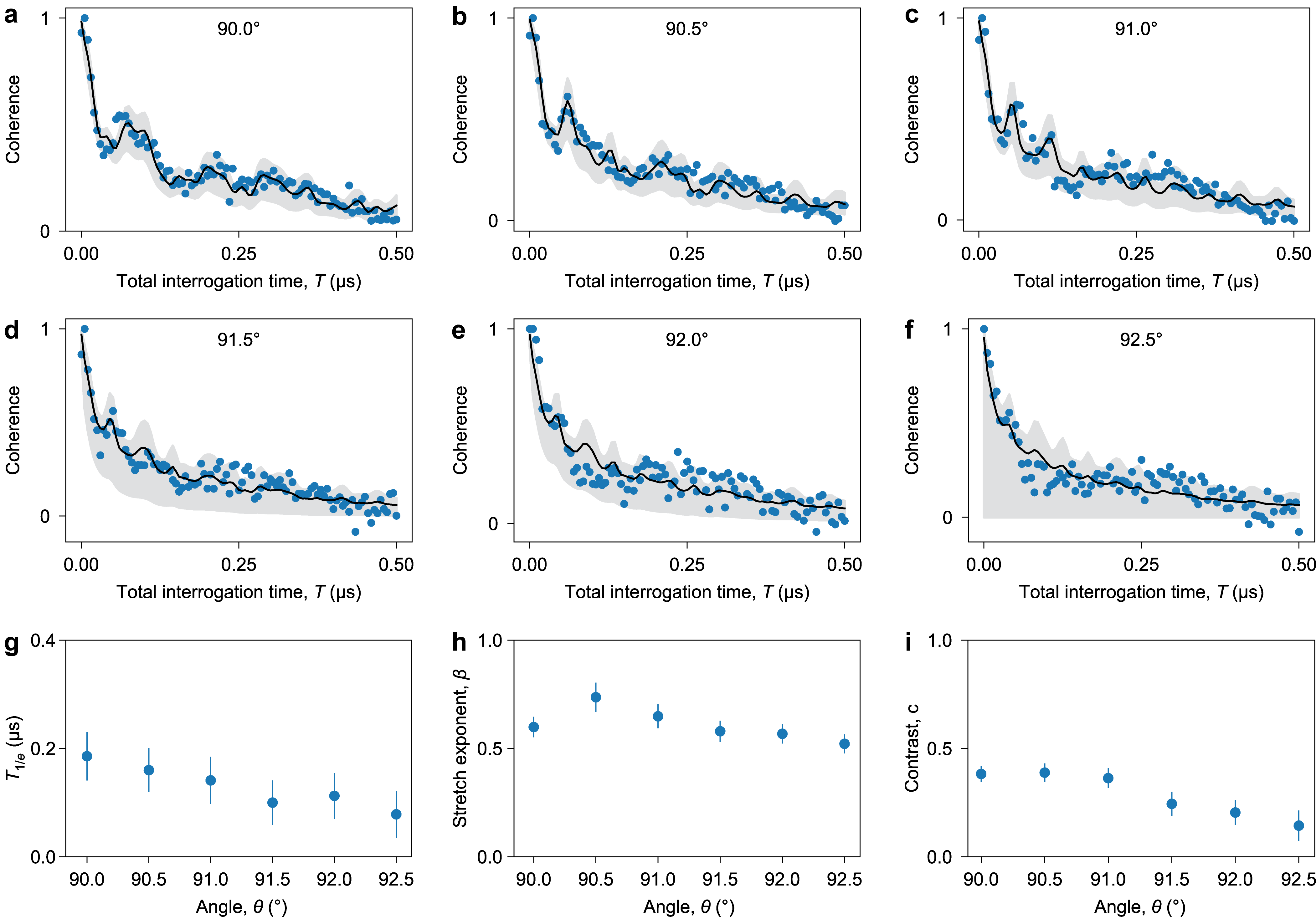}
\supcaption{
\textbf{Angle-resolved spin-echo dynamics under an in-plane magnetic field with corresponding fit analysis.} \textbf{a–f,} Spin-echo coherence modulation data (markers) and fits (solid lines) as a function of the polar angle $\theta$ of the external magnetic field at a fixed amplitude of $20~\mathrm{mT}$. Gray shading denotes the uncertainty of the optimized fit. \textbf{g–i,} Fit parameter dependence on the polar angle $\theta$: (\textbf{g}) $1/e$ decay time, $T_{1/e}$, (\textbf{h}) stretch exponent, $\beta$, and (\textbf{i}) phenomenological contrast, $c$. Error bars indicate fit uncertainties.
}
\end{figure*}

\begin{figure*}[htbp] 
\includegraphics[scale=1]{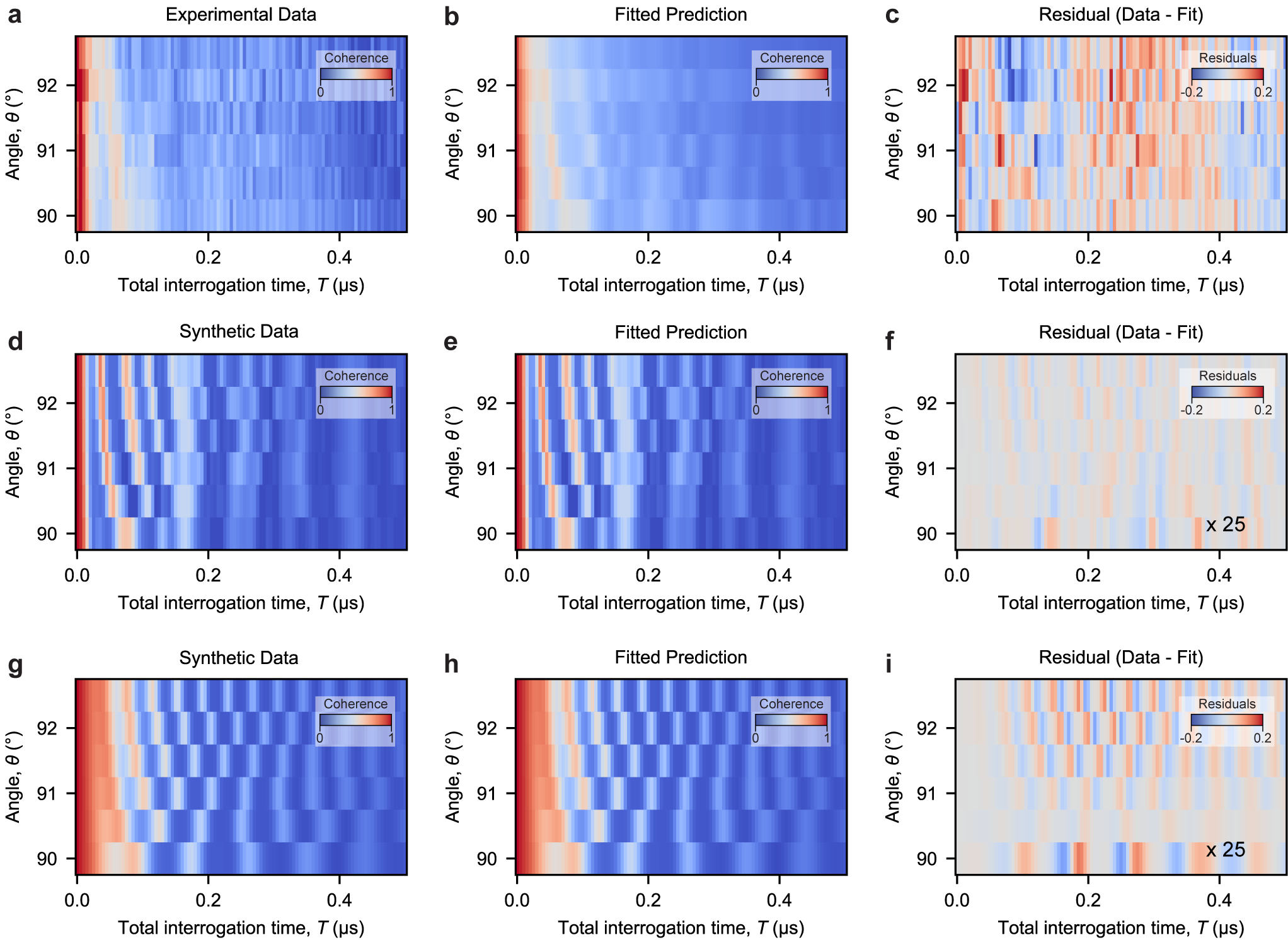}
\supcaption{
\textbf{Validation of our approach for learning hyperfine Hamiltonian parameters.} 
\textbf{a,} Experimental spin-echo coherence dynamics as a function of the polar angle, $\theta$, of the external in-plane magnetic field at a fixed amplitude of $20~\mathrm{mT}$.
\textbf{b,} Predicted spin-echo coherence dynamics based on the hyperfine Hamiltonian parameters summarized in Fig.~2 of the main text.
\textbf{c,} Fit residuals, defined as the difference between the model prediction and the experimental data. To validate our hyperfine parameter learning approach, we numerically generate synthetic data and test whether the same protocol can reliably converge to the ground truth values. \textbf{d,g,} Synthetic spin-echo coherence dynamics, \textbf{e,h,} predicted spin-echo coherence dynamics, and \textbf{f,i,} fit residuals, obtained using \textbf{(d)} the DFT-based hyperfine parameters from Ref.~\cite{ivady2020ab} and \textbf{(g)} the parameters reported in Ref.~\cite{gong2024isotope}. The data is multiplied by 25 for visualization purposes, and displayed in the same colorscale as \textbf(c). As demonstrated by the vanishing residuals in the synthetic data cases, our learning approach reliably reproduces the coherence modulations by recovering the ground truth set of hyperfine interaction parameters.
}
\end{figure*}

\begin{figure*}[htbp]
\includegraphics[scale=1]{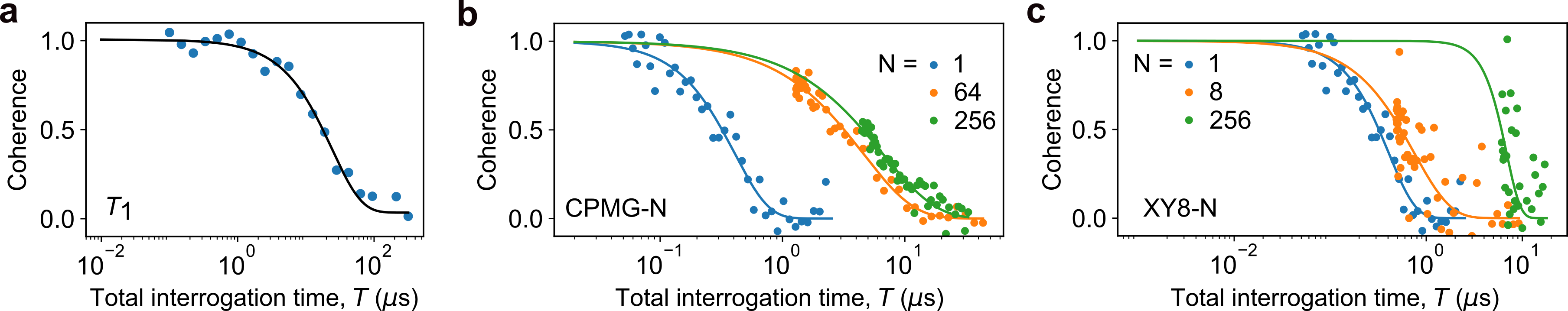}
\supcaption{
\textbf{Probing spin relaxation ($T_1$) and coherence ($T_2$) dynamics at room temperature.}
\textbf{a,} Measurement of the spin relaxation time ($T_{1}$). The $1/e$ $T_1$ time is approximately 10 $\mu$s. \textbf{b,} Coherence profiles measured under CPMG-N for increasing numbers of $\pi$ pulses ($N$). Representative traces are shown for $N = 1$, $64$, and $256$, with $T_{\text{2}}\approx 0.41\pm0.02\units{\mu s}, 4.54\pm0.50\units{\mu s} \text{ and } 7.05\pm1.12\units{\mu s}$, respectively.
\textbf{c,} Coherence profiles measured under XY8-N. Representative traces are shown for $N = 1$, $8$, and $256$, with $T_{\text{2}}\approx 0.41\pm0.02\units{\mu s}, 0.75\pm0.20\units{\mu s} \text{ and } 6.85\pm2.06\units{\mu s}$, respectively.
}
\end{figure*}

\begin{table}[h!]
\centering
\begin{tabular}{lcccc}
\toprule
\multicolumn{5}{c}{Fitted Values from this work} \\
Nuclear Spin& $A_{xx}/2\pi$ (MHz) & $A_{yy}/2\pi$ (MHz) & $A_{zz}/2\pi$ (MHz) & $A_{xy}/2\pi$ (MHz) \\
\midrule
$N_1$ & -104 $\pm 1$   & -34 $\pm 1.3$   & -67 $\pm 0.5$   & 0 \\
$N_2$ & -51 $\pm 0.5$  & -86 $\pm 3.3$   & -67 $\pm 0.5$   & 30 $\pm 1$ \\
$N_3$ & -51 $\pm 0.5$  & -86 $\pm 3.3$   & -67 $\pm 0.5$   & -30 $\pm 1$ \\
\midrule
\multicolumn{5}{c}{Hyperfine values from Ref.~\cite{gong2024isotope}} \\
Nuclear Spin& $A_{xx}/2\pi$ (MHz) & $A_{yy}/2\pi$ (MHz) & $A_{zz}/2\pi$ (MHz) & $A_{xy}/2\pi$ (MHz) \\
\midrule
$N_1$ & -13.720 & -26.322 & -65.900 & 0.000 \\
$N_2$ & -23.208 & -17.001 & -65.900 & 5.375 \\
$N_3$ & -23.208 & -17.001 & -65.900 & -5.366 \\
\midrule
\multicolumn{5}{c}{Hyperfine values from Ref.~\cite{ivady2020ab}} \\
Nuclear Spin& $A_{xx}/2\pi$ (MHz) & $A_{yy}/2\pi$ (MHz) & $A_{zz}/2\pi$ (MHz) & $A_{xy}/2\pi$ (MHz) \\
\midrule
$N_1$ & -65.856 & -126.294 & -65.900 & 0.000 \\
$N_2$ & -111.396  & -81.605 & -65.900 & 25.800 \\
$N_3$ & -111.396  & -81.605 & -65.900 & -25.758 \\
\bottomrule
\end{tabular}
\caption{Hyperfine tensor components and comparison with previous literature.}
\label{tab:hyperfine}
\end{table}

\end{document}